\documentclass[twocolumn,showpacs,superscriptaddress,preprintnumbers,amsmath,amssymb]{revtex4-2}
\usepackage{graphicx}
\usepackage{dcolumn}
\usepackage{appendix}
\usepackage{bm}

\usepackage[usenames,dvipsnames]{xcolor}
\usepackage[caption=false]{subfig}
\usepackage{bbm}
\usepackage{enumerate}

\usepackage{array}
\usepackage{dsfont}
\usepackage{physics}
\usepackage{mathpazo}

\newcommand{\pol}{ {\boldsymbol{\epsilon}} }

\newcommand{\Gtwo}{ G^{(2)} }

\usepackage[
  bookmarks=true,
  colorlinks,
  linkcolor=blue,
  urlcolor=blue,
  citecolor=blue,
  plainpages=false,
  pdfpagelabels,
  final,
  breaklinks=true
]{hyperref}

\begin{document}

\title{Photon anti-bunching in high harmonic generation}

\author{Philipp Stammer}
\email{philipp.stammer@icfo.eu}
\affiliation{ICFO-Institut de Ciencies Fotoniques, The Barcelona Institute of Science and Technology, Castelldefels (Barcelona) 08860, Spain}
\affiliation{Atominstitut, Technische Universit\"{a}t Wien, 1020 Vienna, Austria}

\author{Javier Rivera-Dean}
\affiliation{ICFO-Institut de Ciencies Fotoniques, The Barcelona Institute of Science and Technology, Castelldefels (Barcelona) 08860, Spain}
\affiliation{Department of Physics and Astronomy, University College London, Gower Street, London WC1E 6BT, UK}

\author{Maciej Lewenstein}
\affiliation{ICFO-Institut de Ciencies Fotoniques, The Barcelona Institute of Science and Technology, Castelldefels (Barcelona) 08860, Spain}

\date{\today}

\begin{abstract}

Photon anti-bunching is the direct evidence for the existence of photons without having a classical counterpart. Unlike bunching of photons, which can have a semi-classical description, the effect of photon anti-bunching can only be understood with quantized electromagnetic fields. However, for the process of high harmonic generation (HHG), where many photons of the driving field are upconverted to a single photon of higher energy, there is yet no clear evidence for the presence of individual photon emission. The key result of this work is the prediction of photon anti-bunching in the process of HHG, marking it the first theoretical discovery of non-classicality in the temporal correlations of HHG photons. While other non-classical signatures in HHG, such as sub-Poissonian statistics or squeezing, have been discussed for an ensemble of photons, the anti-bunching signature reported here is a signature of a single photon. This is achieved by using the recently developed Heisenberg picture approach for quantum optical HHG, revealing clear anti-bunching signatures in the intensity correlation function across the entire harmonic spectrum.

\end{abstract}

\maketitle

\begin{figure}
	\includegraphics[width=0.9\columnwidth]{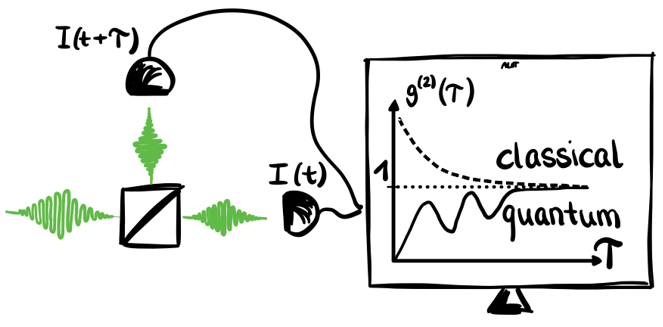}
	\caption{Schematics of an experimental HBT-configuration to measure the second-order field correlation function, $g^{(2)}(\tau)$, by means of the intensity correlations between the detectors at time $I(t)$ and $I(t+\tau)$. While the equal time correlation function $g^{(2)}(0)$ allows to infer on the photon number distribution, only the temporal correlation as a function of $\tau$ can conclude on photon anti-bunching via $g^{(2)}(0) < g^{(2)}(\tau)$.}
      \label{fig:intro}
\end{figure}

The concept of the photon was introduced, among others, by Einstein to explain the photoelectric effect~\cite{einstein1905erzeugung}. Although the notion of the photon persists, the photoelectric effect is no direct evidence for the existence of photons since semi-classical descriptions can explain the phenomena~\cite{lamb1968photoelectric}. 
While many typical quantum optical problems have a semi-classical counterpart, the photon, nevertheless, is a central concept in modern quantum optics~\cite{walls2008quantum, louisell1973quantum, haroche2006exploring}, and its notion is a key element in the quantization of the electromagnetic field~\cite{scully1997quantum}. 
Furthermore, and central for the rise of quantum optics, was the development of the quantum theory of optical coherence~\cite{glauber1963coherent, glauber1963quantum, glauber1963states}, where the wave-particle nature of the electromagnetic field is considered from the perspective of correlation functions~\cite{glauber1963photon}. 
The measurement of field correlation functions encodes the information of the fluctuations of the field, and can give rise to genuine quantum optical signatures without a semi-classical counterpart~\cite{mandel1965coherence, mandel1995optical}. 
This includes the observation of anti-correlated temporal photon detection in the second-order correlation function $g^{(2)}(\tau)$, which is obtained from an intensity correlation measurement between two detectors with time-delay $\tau$ (see Fig.~\ref{fig:intro}). 
The increase of $g^{(2)}(0) < g^{(2)}(\tau)$ for increasing time-delay, indicates the non-classical signature of photon anti-bunching, which was first predicted for the process of resonance fluorescence~\cite{kimble1976theory}, and subsequently measured from the emission of a single~\cite{kimble1977photon} or two atoms~\cite{itano1988photon}. 
While the effect of photon bunching, $g^{(2)}(0) > g^{(2)}(\tau)$, is classical and describes the tendency of the photons to arrive together, the property of anti-bunching is a genuine quantum effect of anti-correlated photon emission. Furthermore, the signature of $g^{(2)}(0) = 0 $ indicates the presence of a single photon~\cite{kimble1977photon}.

At the other end of the photon number scale, there are radiation fields consisting of millions of photons intense enough to induce highly non-linear interaction in media~\cite{brabec2000intense, krausz2009attosecond}. The non-perturbative interaction can lead to the upconversion process of high harmonic generation (HHG), where many photons of the driving field are upconverted to a single harmonic photon, and the sheer number of photons dominated the believe that quantum optical signatures are irrelevant at these intensities, making the semi-classical description of HHG the dominant approach~\cite{lewenstein1994theory, amini2019symphony}.
However, recent advances in the full quantum optical description of HHG has initiated growing interest in finding non-classical signatures in the harmonic radiation~\cite{lewenstein2021generation, gorlach2020quantum, stammer2025colloquium, cruz2024quantum}.
This lead to a variety of theoretical predictions ranging from entanglement between the modes~\cite{stammer2022high, stammer2022theory, yi2024generation, stammer2024entanglement}, to squeezing in the harmonics~\cite{yi2024generation, stammer2024entanglement, lange_excitonic_2025, tzur2024generation}, while first experiments have shown that the exchange of photons can lead to the generation of optical cat states~\cite{lewenstein2021generation, rivera2022strong}, or that the field modes can be entangled~\cite{theidel2024evidence}.
Relevant for this work is the property of the harmonic photon statistics via the equal time correlation function $g^{(2)}(0)$, where previous work on the harmonic photon number distribution has measured classical super-Poissonian photon statistics~\cite{lemieux2024photon}, or predicted a non-classical sub-Poissonian signature~\cite{gorlach2020quantum}. 
However, and despite these achievements, a discussion on the temporal photon correlations in the context of HHG is missing thus far, particularly because the photon number distribution is a physically distinct property than the temporal photon correlations encoded in the full $g^{(2)}(\tau)$ function, where only the latter allows to infer on non-classical anti-bunching~\cite{zou1990photon}.
Furthermore, there is yet no quantum signature of temporal correlations in the $g^{(2)}(\tau)$ function in strong field quantum optics.

In this work, we present the first prediction of such non-classical anti-bunching photon correlations in the process of HHG across the entire spectral range. This is achieved by using the Heisenberg picture approach developed in the companion paper~\cite{stammer2025theory}, allowing to calculate the two-time intensity correlation function.

\emph{Heisenberg picture of HHG.---}
The properties of the fluctuations of the optical field are encoded in correlation functions~\cite{mandel1965coherence}, where the first-order correlation is related to the spectrum of the field, while the signature of photon anti-bunching is encoded in the second-order field correlation.
These are naturally formulated within the Heisenberg picture, where the field operators $a^{(\dagger)}(t)$ are propagated in time. 
For the process of HHG, the Heisenberg picture got little to no attention so far~\cite{sundaram1990high, diestler2008harmonic}, and is re-introduced in the companion paper~\cite{stammer2025theory}, where we formulate the quantum theory of optical coherence for HHG in the Heisenberg picture~\cite{stammer2025theory}.  
Now, considering the typical strong field scenario where an intense laser field, described by the coherent state $\ket{\alpha}$, interacts with an ensemble of $N$ charges in the inital ground state $\ket{\vb{\bar g}} = \otimes_{i=1}^N \ket{g_i}$, the Hamiltonian is $H = H_F + H_S + H_I$.
The field and the interaction Hamiltonian are given by~\cite{stammer2023quantum}
\begin{align}
    H_F = \sum_{q=1}^{q_c} \hbar \omega_q a_q^\dagger a_q, \quad H_I = - \sum_{i=1}^N \vb{d}_i \cdot \vb{E}_Q(t),
\end{align}
where the interaction couples the dipole $\vb{d}_i$ to the electric field operator
\begin{align}
    \vb{E}_Q = - i g\hbar \sum_q \sqrt{q} \, \bm{\epsilon}_q \left( a_q^\dagger - a_q \right).
\end{align}

The system Hamiltonian $H_S$ is generic, and includes the Hamiltonian of the $N$ charges, which could be the single-active electron of gas phase atoms or the charges in a solid state system. While the theory of the Heisenberg picture formulated in Ref.~\cite{stammer2025theory} is valid for all systems, in this work we consider an ensemble of $N$ atoms. 
Given the Hamiltonian $H$, the time-dependent field operator in the Heisenberg picture, $a_q(t) = U^\dagger(t) a_q(0) U(t)$, can be solved explicitly (see Appendix~\ref{end:heisenberg})
\begin{equation}
\begin{aligned}
\label{eq:aq_heisenberg_solution}
    a_{q}(t) & =  a_{q} \,  e^{- i \omega_{q} t} \\
    & \quad + g \sqrt{q} \int_{t_0}^t \dd t^\prime e^{- i \omega_{q} (t-t^\prime)} \sum_{i=1}^N \pol_{q} \cdot \vb{d}_i(t^\prime).
\end{aligned}
\end{equation}

The dipole moment $\vb{d}(t) = U^\dagger(t) \vb{d}(0) U(t)$, acting as the source term for the scattered field, is obtained from the respective Heisenberg equation of motion~\cite{stammer2025theory}. 
Given the experimental situation of interest, where the driving field has typical values of $\abs{\alpha} \sim 10^6$, we can consider the dipole to be driven to zeroth order in the quantum optical field, such that $\vb{\tilde d}(t)= U_{sc}^\dagger(t) \vb{d}(0) U_{sc}(t)$. Hence, the dipole operator is propagated by the semi-classical Hamiltonian $H_{sc}(t) = H_S + H_{I,cl}(t)$, where the classical interaction is $H_{I,cl}(t) = - \sum_i \vb{d}_i \cdot \vb{E}_{cl}(t)$, and the classical field is obtained from $\vb{E}_{cl}(t) = \Tr[\vb{E}_Q(t) \dyad{\alpha}]$.
In the companion paper~\cite{stammer2025theory}, we have provided the detailed derivation together with an analysis of the first-order field correlation function $G^{(1)}(t_1,t_2) = \expval*{E^{(-)}(t_1) E^{(+)}(t_2)}$ and corresponding spectrum $S(\omega)$. Here, $E^{(+)}(t) = i \hbar g \sum_q \sqrt{q} \bm{\epsilon}_q a_q(t)$ is the positive component of the field operator with $E^{(-)}(t) = [E^{(+)}(t)]^\dagger$.
It was shown that HHG in general exhibits first-order coherence, $\abs*{G^{(1)}(t_1, t_2)} = [G^{(1)}(t_1,t_1) G^{(1)}(t_2,t_2) ]^{1/2} $, and how the HHG spectrum is obtained when accounting for the probabilistic nature of the dipole by using the Wiener-Khintchine theorem~\cite{wiener1930generalized, khintchine1934korrelationstheorie}, naturally leading to a coherent and incoherent HHG spectrum~\cite{stammer2025theory, gorlach2020quantum, stammer2025quantum}.

\emph{Second-order correlation function in HHG.---}
In this work, we extend the quantum theory of optical coherence to the second-order field correlation function
\begin{align}
\label{eq:correlation_2nd_order_general}
    G^{(2)}(t_1,t_2) & = \expval*{E^{(-)}(t_1)E^{(-)}(t_2)E^{(+)}(t_2)E^{(+)}(t_1)},
\end{align}
which can be re-written in terms of the normally ordered intensity correlation $ G^{(2)}(t_1,t_2) = \expval*{: I(t_1) I(t_2) :}$. 
Hence, the second-order correlation function is essentially an intensity correlation measurement, where the experiment measures the joint photocurrent of photon detection at time $t_1=t$, and time $t_2 = t+\tau$. Such a configuration was introduced by Hanbury Brown and Twiss (HBT), showing photon bunching for chaotic light~\cite{brown1956correlation}. Such a HBT-configuration for an intensity correlation measurement is shown in Fig.~\ref{fig:intro}, where the normalized intensity correlation 
\begin{align}
    g^{(2)}(\tau) & = \frac{\expval*{: I(t)I(t+\tau) :}}{\expval*{I(t)} \expval*{I(t+\tau)}},
\end{align}
is non-classical only for $g^{(2)}(\tau) < 1$, and $g^{(2)}(\tau) \geq 1$ correspond to classical signatures~\cite{mandel1966photon, mandel1995optical}.

We are now interested in obtaining the intensity correlation function of the harmonics by correlating the measured intensities from two detectors with a time-delay $\tau$.
Using the second-order correlation function from Eq.~\eqref{eq:correlation_2nd_order_general} we can write 
\begin{align}
\label{eq:gtwo_harmonic_q}
    \Gtwo (t,t+\tau) = \hbar^4 g^4 q^2 \expval{\mathcal{N}(t,\tau)},
\end{align}
where we have defined the normal-ordered expectation value of the harmonic field operators $\expval*{\mathcal{N}(t,\tau)} = \expval*{a_q^\dagger(t) a_q^\dagger (t+\tau) a_q(t+\tau) a_q(t)}$.
Taking into account that the harmonics are initially in the vacuum $\ket*{0_q}$, we only consider the contribution of the scattered field in Eq.~\eqref{eq:aq_heisenberg_solution} (second term), since the free-field contribution vanishes in the vacuum (first term).
For the two-time average of the normal ordered field operators, we find
\begin{equation}
\begin{aligned}
\label{eq:expectation_value_dipoles}
    \expval{\mathcal{N}(t,\tau)} &= g^4 q^2 \int_{t_0}^{t} dt_1 \int_{t_0}^{t+\tau} dt_2 \int_{t_0}^{t+\tau} dt_3 \int_{t_0}^{t} dt_4 \\
    & \quad\times e^{- i \omega_q (t_1+t_2-t_3 -t_4)} \mathcal{D}(t_1,t_2,t_3,t_4),
\end{aligned}
\end{equation}
where we have defined the dipole correlation term
\begin{equation}
\begin{aligned}
\label{eq:dipole_correlation_2nd_order}
    \mathcal{D}(t_1,t_2,t_3,t_4) & \equiv \sum_{i,j,k,l}^N \Tr \left[ (\pol_q \cdot \vb{d}_i (t_1)) (\pol_q \cdot \vb{d}_j (t_2)) \right. \\
    & \left. \quad \times (\pol_q \cdot \vb{d}_k (t_3)) (\pol_q \cdot \vb{d}_l (t_4)) \rho_S(t_0) \right],
\end{aligned}
\end{equation}
where $\rho_S(t_0)$ is the initial state of the system. 
If we only consider the dipole along the laser polarization direction, $\pol_q \cdot \vb{d}_i(t) \equiv d_i(t)$, within the $N=1$ single atom picture, we now have
\begin{align}
\label{eq:dipole_4th_order}
     \mathcal{D}(\boldsymbol{t} ) & = \bra{\text{g}} d(t_1) d(t_2) d(t_3) d(t_4) \ket{\text{g}},
\end{align}
where we have used $\rho_S(t_0) = \dyad{g}$, and introduced the tuple over the different time arguments $\boldsymbol{t} \equiv (t_1, t_2, t_3,t_4)$.
Using that the second-order dipole correlation can be decomposed as $\expval{d(t_1) d(t_2)} = \expval{d(t_1)} \expval{d(t_2)} + \expval{\Delta d(t_1) \Delta d(t_2)}$, with the fluctuations around the mean $\Delta d(t) = d(t) - \expval{d(t)}$, we can write for the 4th-order dipole correlation
\begin{equation}
\begin{aligned}
    \mathcal{D}(\boldsymbol{t}) = & \, \expval{d(t_1) d(t_2)} \expval{d(t_3) d(t_4)} \\
    & + \expval{\Delta [d(t_1) d(t_2)] \Delta [d(t_3) d(t_4)]}, 
\end{aligned}
\end{equation}
where we have defined 
\begin{align}
    \Delta [d(t_1) d(t_2)] \equiv d(t_1) d(t_2) - \expval{d(t_1) d(t_2)}.
\end{align}

\begin{figure*}
    \centering
	\includegraphics[width = \textwidth]{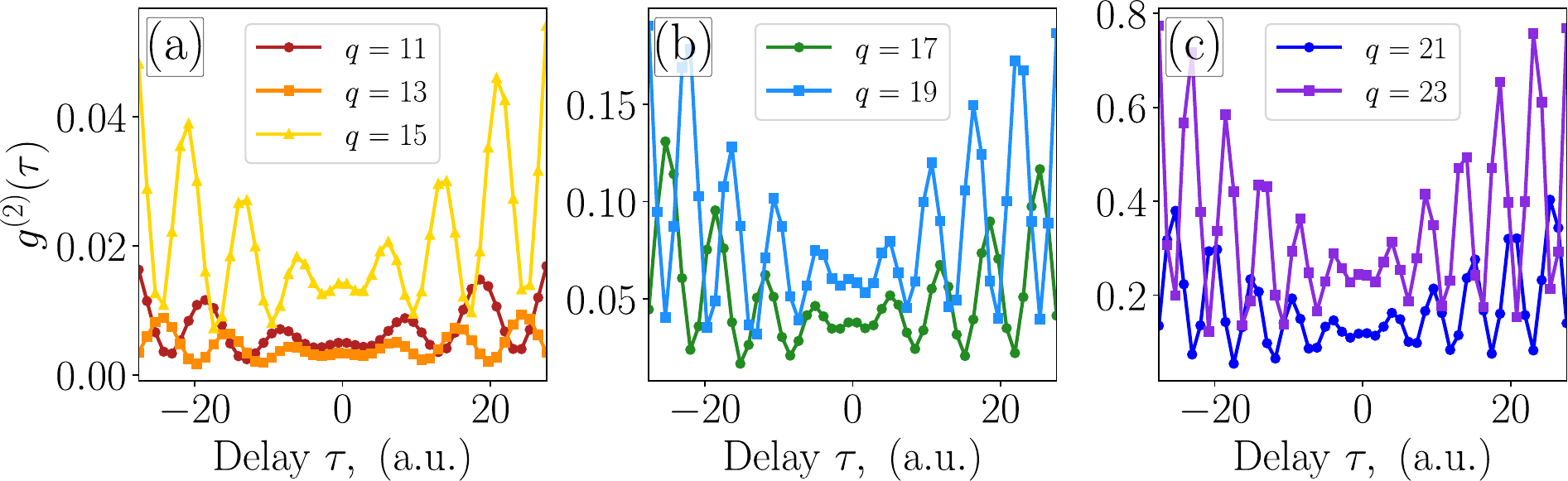}
	\caption{We show the normalized second-order correlation function $g^{(2)}(\tau)$ for the relevant range of non-perturbative odd harmonics in the plateau (a) $q \in \{11,\, 13,\, 15 \}$  and (b) $q \in \{17, \, 19\}$, while in (c) $q \in \{21, \, 23 \}$  near the cut-off (for the low-order harmonics we refer to the Appendix~\ref{end:low_order} in which the same behavior is shown). The time-delay $\tau$ is taken to be after one complete cycle of the driving laser field $T = 2 \pi / \omega$, and evaluated at $K=50$ delay times. The driving field parameters are $E_0 = 0.053$ a.u. $\omega_L = 0.057$ a.u., and $I_p = 0.5$ a.u., corresponding to that of hydrogen (see SM for details).}
      \label{fig:g2_function}
\end{figure*}

We can see that the 4th-order dipole correlation is composed out of the product of two dipole correlation terms, and a term including the fluctuations of the dipole correlations $\Delta [d(t_1) d(t_2)]$.
Now, equipped with the expression for the second-order correlation function, we can look at the normalized correlation function of a given harmonic 
\begin{align}
\label{eq:g2_definition_field_operators}
    g^{(2)}(\tau) & = \frac{\expval*{:I_q(t) I_q(t+\tau):}}{\expval*{I_q(t)} \expval*{I_q(t+\tau)}}.
\end{align}

This can in fact be written in terms of the first and second-order correlation function
\begin{align}
    g^{(2)}(\tau) = \frac{G^{(2)}(t,t+\tau)}{G^{(1)}(t+\tau) G^{(1)}(t)},
\end{align}
and we can use the results from the companion paper~\cite{stammer2025theory}, that the first-order correlation function in the denominator can be written as (see Appendix~\ref{end:heisenberg} for details)
\begin{align}
    G^{(1)} (t) = G^{(1)}_{coh}(t) + G^{(1)}_{inc}(t). 
\end{align}

Taking into account that there is no commutator structure in classical intensity correlation measurements, it is known that, $g^{(2)}_{cl}(0) \geq g^{(2)}_{cl}(\tau)$ and $g^{(2)}_{cl}(0) \geq 1$, must hold~\cite{mandel1995optical}. A violation of these inequalities is therefore a direct witness of non-classical temporal photon correlation properties. 
Now, with the expression of the 4th-order dipole correlation in Eq.~\eqref{eq:dipole_4th_order}, for which a more detailed derivation is given in the Supplementary Material (SM), we can now compute the $g^{(2)}$-function for varying time-delay $\tau$. We solve the second-order correlation function $G^{(2)}(t,t+\tau)$ numerically, such that all orders of the dipole moment correlations are taken into account. 
The normalized intensity correlations $g^{(2)}(\tau)$ emitted from one cycle of the driving field $t = T = 2 \pi / \omega$, are shown in Fig.~\ref{fig:g2_function}~(a)-(c) for all higher harmonic orders $q \in [11,23]$. 
Comparing the HHG intensity correlation in Fig.~\ref{fig:g2_function} with the classical constraints, we make the first important discovery, and one key result of this Letter, namely that the high harmonic $g^{(2)}(\tau)$ function violates both classical inequalities. First, via $g^{(2)}(0) < 1$, and second the increase of the correlation function compared to the correlations at vanishing time-delay $g^{(2)}(0) < g^{(2)}(\tau)$.
This provides the first quantum signature of high harmonic emission in the temporal photon correlations, showing the property of photon anti-bunching across all harmonic orders. For vanishing time delay, $g^{(2)}(0)$, the values are close to that of perfect single photon emission. For instance, we have $g^{(2)}_{11}(0) \approx 5\times 10^{-3}$, and $g^{(2)}_{13}(0) \approx 3\times10^{-3}$, for the harmonics 11 and 13, respectively.  
This constitutes the only discussion of temporal photon correlations in strong field quantum optics via the full $g^{(2)}(\tau)$, predicting non-classical signatures.
Furthermore, we can see a decrease in the anti-correlation measure for increasing time-delay (increase of the $g^{(2)}(\tau)$ function over time). This indicates the expected property that for large $\tau$ the photon correlations decrease, and will eventually become uncorrelated ($g^{(2)} = 1$) in the long time limit.

The phenomena of photon anti-bunching in HHG can intuitively be understood from the powerful picture of the 3-step model and the corresponding electron trajectories~\cite{corkum1993plasma, lewenstein1994theory, salieres_feynmans_2001}.
Once an electron is ionized and accelerated in the continuum, it emits the harmonic photon upon recombination with the ion. Then, in order to emit another photon, the electron needs to undergo the whole 3-steps of ionization, propagation and recombination for the next emission event. Since this process is not instantaneous, the electron does not emit two photons at the same time, leading to anti-bunching signatures.  
It implies that there are hardly ever two photons emitted at the same time, independent of the driving intensity.
We therefore emphasize that the explanation using the 3-step model shows how fundamentally different HHG is from the general perturbative non-linear optical response. 
However, note that the 3-step model is by itself not a quantum optical theory, but provides an intuitive picture for understanding the anti-bunching phenomena, and shares the same intuition with the common explanation of anti-bunching in resonance fluorescence~\cite{kimble1977photon, kimble1976theory}. In this case, after emission of a photon by a single atom, it can not emit a second photon before the atom gets excited again~\cite{carmichael2013statistical}. Nevertheless, further analysis into the mechanism of this novel anti-bunching signature in HHG seems promising to uncover the full quantum optical emission dynamics of HHG photons on the attosecond time scale.
Furthermore, it is important to note that the presented intensity correlation function and associated anti-bunching, $g^{(2)}(0) < g^{(2)}(\tau)$, is shown on the femtosecond time-scale, implying that standard measuring techniques using photodetectors reach their limitations, and that alternative schemes using photon absorption techniques with femtosecond resolution of the $g^{(2)}(\tau)$ function are needed~\cite{boitier2009measuring}. However, resolving the attosecond oscillations in $g^{(2)}(\tau)$ would require novel measurement schemes, which will hopefully inspire further investigation into combining attosecond techniques with quantum optical measurements~\cite{rivera2026attosecond, lamprou2025nonlinear, tzur2025measuring, stammer2025weak}.

\emph{Many atom picture.---}
So far, we have studied the single atom scenario, and therefore only a single emitter will contribute to the emission of harmonics and intensity correlation function. Akin to the first-order correlation function discussed in Ref.~\cite{stammer2025theory}, we are now interested in the many atom case. 
In the following we will consider the case of $N$ indistinguishable emitters, such that the higher order dipole correlation in Eq.~\eqref{eq:dipole_4th_order} along the laser polarization direction is given by (details can be found in the Appendix~\ref{end:many_atom})
\begin{align}
\label{eq:dipole_high_order_expanded}
    &\sum_{i,j,k,l=1}^N  \bra{\vb{\bar g}} d_i(t_1) d_j(t_2) d_k(t_3) d_l(t_4) \ket{\vb{\bar g}}  \\
    & = \frac{N!}{(N-4)!} \expval{d(t_1)} \expval{d(t_2)} \expval{d(t_3)} \expval{d(t_4)} + \mathcal{O}(N^3). \nonumber
\end{align}

Interestingly, the coherent contribution is most dominant in the higher order correlation functions when considering the many atom regime, with a $\mathcal{O}(N^4)$ scaling. 
While typical HHG experiments are conducted in the regime of many atoms, it promotes further investigation to uncover the non-classical emission characteristics in HHG. This includes sub-Poissonian photon statistics~\cite{gorlach2020quantum}, and the predicted anti-bunching of this work. 
Since the analysis of this work considered uncorrelated emitters, and it has already been shown that the emission characteristics from correlated systems can differ~\cite{pizzi2023light}, collective quantum jumps seems to be a promising direction~\cite{lewenstein1987cooperative, itano1988photon}.

\begin{figure}
    \centering
	\includegraphics[width = \columnwidth]{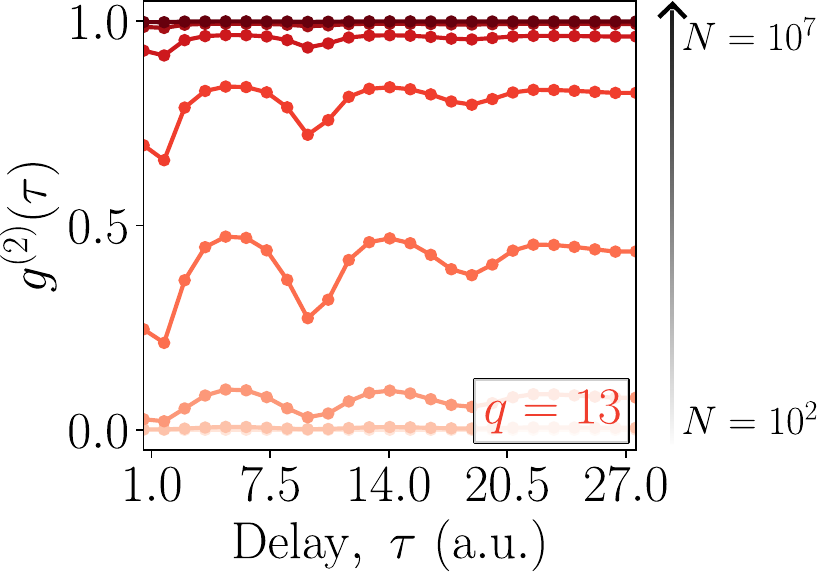}
	\caption{Normalized intensity correlation $g^{(2)}(\tau)$ of the $q= 13$ harmonic order for varying number of emitters $N$ (harmonics $q=19$ and $q=23$ are shown in Fig.~\ref{fig:g2_endmatter} of the Appendix~\ref{end:low_order}). The atom number increases for increasing color brightness including $\log_{10}N \in \{2.0, 2.7,3.4,4.1,4.8,5.5,6.7,7.0 \} $. The same field conditions as those in Fig.~\ref{fig:g2_function} are considered.}
      \label{fig:dipole_higher_order}
\end{figure}

In order to discuss the influence of the many atom emission on the intensity correlation, we show in Fig.~\ref{fig:dipole_higher_order} the $g^{(2)}(\tau)$ function for different values for the number of atoms $N$. As expected, the second-order correlation function increases for increasing number of uncorrelated emitters until it approaches unity for very large $N$. This is attributed to the different scalings of the coherent and incoherent contributions, where the former has the $N^4$ scaling whereas the latter scales at most with $\mathcal{O}(N^3)$. The limit of $g^{(2)}(\tau) = 1$ is the well known signature of classical coherent radiation due to oscillating charge currents, here given by the dipole moment expectation value of the coherent contribution.

\begin{figure}
	\includegraphics[width = 1.0\columnwidth]{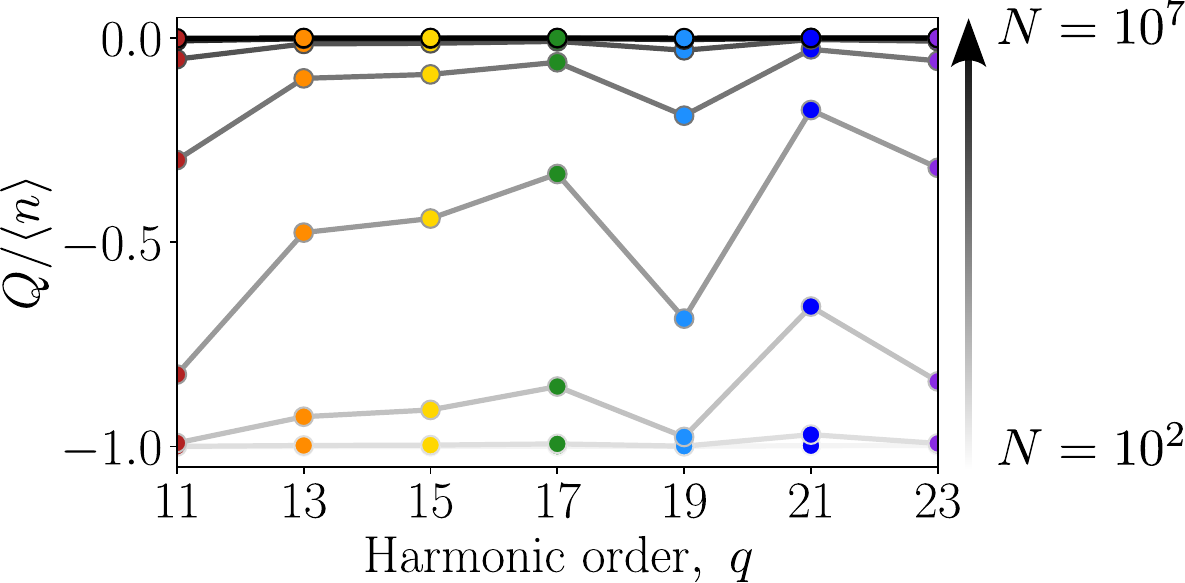}
	\caption{Mandel $Q$-parameter as a function of the harmonic orders indicating sub-Poissonian photon statistics as $Q < 0$. The $Q$-parameter is shown for increasing number of emitters $N$ including $\log_{10}N \in \{2.0, 2.7,3.4,4.1,4.8,5.5,6.7,7.0 \} $, where the guiding line indicates a fixed emitter number, and the colors are those from Fig.~\ref{fig:g2_function} for the respective harmonic order. The same field conditions as those in Fig.~\ref{fig:g2_function} are used.}
      \label{fig:mandel}
\end{figure}

\emph{Photon statistics in quantum HHG.---}
So far we have presented the intensity correlation function in HHG, which we will now use to discuss the related measure of the photon number distribution of the field, encoded in the Mandel $Q$-parameter~\cite{mandel1979sub}. The $Q$-parameter is a measure for the deviation of the photon number distribution from that of a Poisson process. The Poisson distribution ($Q=0$) is obtained from coherent states in which the variance of the distribution is equal to the mean. The case of $Q<0$, and $Q>0$, corresponds to sub-Poissonian and super-Poissonian distributions, respectively. Here, only the sub-Poissonian statistics corresponds to non-classical signatures in which the variance of the photon number distribution is smaller than its mean (and vice versa for super-Poissonian distributions). 
Given the results obtained for the second-order correlation function, showing photon anti-bunching in Fig.~\ref{fig:g2_function}, we can infer that the photon number distribution of the harmonic field modes exhibit non-classical sub-Poissonian statistics. This is due to the relation $Q = \expval*{n} (g^{(2)}(0) - 1)$, and hence $Q < 0$ due to $g^{(2)} (0) < 1$. We therefore show the Mandel $Q$-parameter as a function of the harmonic order in Fig.~\ref{fig:mandel}, showing the aforementioned sub-Poissonian photon number distribution -- a further indicator of non-classicality in the field. This is in agreement with the findings in Ref.~\cite{gorlach2020quantum}, where negative $Q$ parameters where shown for HHG.

Similarly to above, where we have discussed the property of the $g^{(2)}$-function with respect to the number of emitters $N$, we also show the dependence of the photon number distribution on the number of emitters. Given the scaling with respect to the number of emitters $N$, we find non-classical sub-Poissonian photon number distributions for an intermediate number of emitters as shown in Fig.~\ref{fig:mandel}. Only in the thermodynamic limit of many emitters, we find that the $Q$-parameter approaches zero, corresponding to the Poisson distribution. This is in contrast to the thermal like photon number distribution reported in Ref.~\cite{gorlach2020quantum}, where the positive $Q$-parameter indicates super-Poissonian statistics.

To conclude the discussion on the photon number distribution, we emphasize that the property of photon anti-bunching and sub-Poissonian statistics share common signatures, but are ultimately different~\cite{zou1990photon}. In this work, we have shown both, anti-bunching in the emitted harmonics as the first quantum signature of temporal correlations in the photon emission process for HHG, as well as the signature of sub-Poissonian statistics in the distribution of the photon number.

\emph{Discussion.---}
In conclusion, we have predicted the first direct non-classical photon signature in the HHG emission process by means of anti-bunching in the temporal photon correlations, and is found for the entire harmonic spectrum. 
The anti-bunching signature is clearly visible in the single atom emission, but also persists in the few-body regime. Only when the atom number is significantly large, the intensity correlation shows a Poissonian property of random photon measurements. 
This raises the interest for conducting HHG experiments in the single or few-body emitter regime, for instance in ultracold quantum gases~\cite{stammer2025high} or using quantum simulators~\cite{arguello2024analog, arguello2025quantum}, to observe the quantum nature of the temporal emission statistics. This would ultimately allow to gain further insights into the emission dynamics of HHG photons from quantum jumps~\cite{bergquist1986observation}, which can reveal anti-bunching signatures in the emission from more than one atom~\cite{itano1988photon}, with particular interest in the statistics from collective effects such as correlated quantum jumps~\cite{lewenstein2002quantum, lewenstein1987cooperative}.
The importance of the findings goes beyond the fundamental insights about the quantum nature of the HHG photon emission process. The phenomena of anti-bunching itself is of significant interest because it is a deterministic non-event.
Finally, we emphasize on the importance to distinguish between photon anti-bunching, reported for the first time in this work for HHG, and sub-Poissonian statistics (as reported in Ref.~\cite{gorlach2020quantum}). Although both notions are related, they are not the same~\cite{zou1990photon}. Since sub-Poissonian photon counting does not need to imply photon anti-bunching, we stress that these two phenomena should not be confused.
Hence, this work is the first presentation of non-classical photon correlations in their temporal statistics.

\begin{acknowledgments}

P.S. would like to express his gratitude to Alexander Carmele for the introduction to quantum optical methods and the perspective of the observer.
P.S. acknowledges funding from the European Union’s Horizon 2020 research and innovation programe under the Marie Skłodowska-Curie grant agreement No 847517.~ICFO group acknowledges support from: Ministerio de Ciencia y Innovation Agencia Estatal de Investigaciones (R$\&$D project CEX2019-000910-S, AEI/10.13039/501100011033, Plan National FIDEUA PID2019-106901GB-I00, FPI), Fundació Privada Cellex, Fundació Mir-Puig, and from Generalitat de Catalunya (AGAUR Grant No. 2017 SGR 1341, CERCA program), and MICIIN with funding from European Union NextGenerationEU(PRTR-C17.I1) and by Generalitat de Catalunya and EU Horizon 2020 FET-OPEN OPTOlogic (Grant No 899794) and ERC AdG NOQIA.

\end{acknowledgments}

\bibliography{literatur}{}

\begin{thebibliography}{58}%
\makeatletter
\providecommand \@ifxundefined [1]{%
 \@ifx{#1\undefined}
}%
\providecommand \@ifnum [1]{%
 \ifnum #1\expandafter \@firstoftwo
 \else \expandafter \@secondoftwo
 \fi
}%
\providecommand \@ifx [1]{%
 \ifx #1\expandafter \@firstoftwo
 \else \expandafter \@secondoftwo
 \fi
}%
\providecommand \natexlab [1]{#1}%
\providecommand \enquote  [1]{``#1''}%
\providecommand \bibnamefont  [1]{#1}%
\providecommand \bibfnamefont [1]{#1}%
\providecommand \citenamefont [1]{#1}%
\providecommand \href@noop [0]{\@secondoftwo}%
\providecommand \href [0]{\begingroup \@sanitize@url \@href}%
\providecommand \@href[1]{\@@startlink{#1}\@@href}%
\providecommand \@@href[1]{\endgroup#1\@@endlink}%
\providecommand \@sanitize@url [0]{\catcode `\\12\catcode `\$12\catcode `\&12\catcode `\#12\catcode `\^12\catcode `\_12\catcode `\%12\relax}%
\providecommand \@@startlink[1]{}%
\providecommand \@@endlink[0]{}%
\providecommand \url  [0]{\begingroup\@sanitize@url \@url }%
\providecommand \@url [1]{\endgroup\@href {#1}{\urlprefix }}%
\providecommand \urlprefix  [0]{URL }%
\providecommand \Eprint [0]{\href }%
\providecommand \doibase [0]{https://doi.org/}%
\providecommand \selectlanguage [0]{\@gobble}%
\providecommand \bibinfo  [0]{\@secondoftwo}%
\providecommand \bibfield  [0]{\@secondoftwo}%
\providecommand \translation [1]{[#1]}%
\providecommand \BibitemOpen [0]{}%
\providecommand \bibitemStop [0]{}%
\providecommand \bibitemNoStop [0]{.\EOS\space}%
\providecommand \EOS [0]{\spacefactor3000\relax}%
\providecommand \BibitemShut  [1]{\csname bibitem#1\endcsname}%
\let\auto@bib@innerbib\@empty
\bibitem [{\citenamefont {Einstein}(1905)}]{einstein1905erzeugung}%
  \BibitemOpen
  \bibfield  {author} {\bibinfo {author} {\bibfnamefont {A.}~\bibnamefont {Einstein}},\ }\bibfield  {title} {\bibinfo {title} {{\"U}ber einem die erzeugung und verwandlung des lichtes betreffenden heuristischen gesichtspunkt},\ }\href {https://onlinelibrary.wiley.com/doi/10.1002/andp.19053220607} {\bibfield  {journal} {\bibinfo  {journal} {Annalen der physik}\ }\textbf {\bibinfo {volume} {4}} (\bibinfo {year} {1905})}\BibitemShut {NoStop}%
\bibitem [{\citenamefont {Lamb~Jr}\ and\ \citenamefont {Scully}(1968)}]{lamb1968photoelectric}%
  \BibitemOpen
  \bibfield  {author} {\bibinfo {author} {\bibfnamefont {W.~E.}\ \bibnamefont {Lamb~Jr}}\ and\ \bibinfo {author} {\bibfnamefont {M.~O.}\ \bibnamefont {Scully}},\ }\href {https://ntrs.nasa.gov/api/citations/19680009569/downloads/19680009569.pdf} {\emph {\bibinfo {title} {The photoelectric effect without photons}}},\ \bibinfo {type} {Tech. Rep.}\ (\bibinfo {year} {1968})\BibitemShut {NoStop}%
\bibitem [{\citenamefont {Walls}\ and\ \citenamefont {Milburn}(2008)}]{walls2008quantum}%
  \BibitemOpen
  \bibfield  {author} {\bibinfo {author} {\bibfnamefont {D.~F.}\ \bibnamefont {Walls}}\ and\ \bibinfo {author} {\bibfnamefont {G.~J.}\ \bibnamefont {Milburn}},\ }\href@noop {} {\emph {\bibinfo {title} {Quantum optics}}}\ (\bibinfo  {publisher} {Springer Science \& Business Media},\ \bibinfo {year} {2008})\BibitemShut {NoStop}%
\bibitem [{\citenamefont {Louisell}(1973)}]{louisell1973quantum}%
  \BibitemOpen
  \bibfield  {author} {\bibinfo {author} {\bibfnamefont {W.~H.}\ \bibnamefont {Louisell}},\ }\href@noop {} {\emph {\bibinfo {title} {Quantum statistical properties of radiation}}}\ (\bibinfo  {publisher} {John Wiley and Sons, Inc., New York},\ \bibinfo {year} {1973})\BibitemShut {NoStop}%
\bibitem [{\citenamefont {Haroche}\ and\ \citenamefont {Raimond}(2006)}]{haroche2006exploring}%
  \BibitemOpen
  \bibfield  {author} {\bibinfo {author} {\bibfnamefont {S.}~\bibnamefont {Haroche}}\ and\ \bibinfo {author} {\bibfnamefont {J.-M.}\ \bibnamefont {Raimond}},\ }\href@noop {} {\emph {\bibinfo {title} {Exploring the quantum: atoms, cavities, and photons}}}\ (\bibinfo  {publisher} {Oxford university press},\ \bibinfo {year} {2006})\BibitemShut {NoStop}%
\bibitem [{\citenamefont {Scully}\ and\ \citenamefont {Zubairy}(1997)}]{scully1997quantum}%
  \BibitemOpen
  \bibfield  {author} {\bibinfo {author} {\bibfnamefont {M.~O.}\ \bibnamefont {Scully}}\ and\ \bibinfo {author} {\bibfnamefont {M.~S.}\ \bibnamefont {Zubairy}},\ }\href@noop {} {\emph {\bibinfo {title} {Quantum optics}}}\ (\bibinfo  {publisher} {Cambridge university press},\ \bibinfo {year} {1997})\BibitemShut {NoStop}%
\bibitem [{\citenamefont {Glauber}(1963{\natexlab{a}})}]{glauber1963coherent}%
  \BibitemOpen
  \bibfield  {author} {\bibinfo {author} {\bibfnamefont {R.~J.}\ \bibnamefont {Glauber}},\ }\bibfield  {title} {\bibinfo {title} {Coherent and incoherent states of the radiation field},\ }\href {https://journals.aps.org/pr/abstract/10.1103/PhysRev.131.2766} {\bibfield  {journal} {\bibinfo  {journal} {Phys. Rev.}\ }\textbf {\bibinfo {volume} {131}},\ \bibinfo {pages} {2766} (\bibinfo {year} {1963}{\natexlab{a}})}\BibitemShut {NoStop}%
\bibitem [{\citenamefont {Glauber}(1963{\natexlab{b}})}]{glauber1963quantum}%
  \BibitemOpen
  \bibfield  {author} {\bibinfo {author} {\bibfnamefont {R.~J.}\ \bibnamefont {Glauber}},\ }\bibfield  {title} {\bibinfo {title} {The quantum theory of optical coherence},\ }\href {https://journals.aps.org/pr/abstract/10.1103/PhysRev.130.2529} {\bibfield  {journal} {\bibinfo  {journal} {Phys. Rev.}\ }\textbf {\bibinfo {volume} {130}},\ \bibinfo {pages} {2529} (\bibinfo {year} {1963}{\natexlab{b}})}\BibitemShut {NoStop}%
\bibitem [{\citenamefont {Glauber}(1963{\natexlab{c}})}]{glauber1963states}%
  \BibitemOpen
  \bibfield  {author} {\bibinfo {author} {\bibfnamefont {R.~J.}\ \bibnamefont {Glauber}},\ }\bibfield  {title} {\bibinfo {title} {Coherent and incoherent states of the radiation field},\ }\href {https://doi.org/10.1103/PhysRev.131.2766} {\bibfield  {journal} {\bibinfo  {journal} {Physical Review}\ }\textbf {\bibinfo {volume} {131}},\ \bibinfo {pages} {2766} (\bibinfo {year} {1963}{\natexlab{c}})}\BibitemShut {NoStop}%
\bibitem [{\citenamefont {Glauber}(1963{\natexlab{d}})}]{glauber1963photon}%
  \BibitemOpen
  \bibfield  {author} {\bibinfo {author} {\bibfnamefont {R.~J.}\ \bibnamefont {Glauber}},\ }\bibfield  {title} {\bibinfo {title} {Photon correlations},\ }\href {https://doi.org/10.1103/PhysRevLett.10.84} {\bibfield  {journal} {\bibinfo  {journal} {Physical Review Letters}\ }\textbf {\bibinfo {volume} {10}},\ \bibinfo {pages} {84} (\bibinfo {year} {1963}{\natexlab{d}})}\BibitemShut {NoStop}%
\bibitem [{\citenamefont {Mandel}\ and\ \citenamefont {Wolf}(1965)}]{mandel1965coherence}%
  \BibitemOpen
  \bibfield  {author} {\bibinfo {author} {\bibfnamefont {L.}~\bibnamefont {Mandel}}\ and\ \bibinfo {author} {\bibfnamefont {E.}~\bibnamefont {Wolf}},\ }\bibfield  {title} {\bibinfo {title} {Coherence properties of optical fields},\ }\href {https://journals.aps.org/rmp/abstract/10.1103/RevModPhys.37.231} {\bibfield  {journal} {\bibinfo  {journal} {Reviews of modern physics}\ }\textbf {\bibinfo {volume} {37}},\ \bibinfo {pages} {231} (\bibinfo {year} {1965})}\BibitemShut {NoStop}%
\bibitem [{\citenamefont {Mandel}\ and\ \citenamefont {Wolf}(1995)}]{mandel1995optical}%
  \BibitemOpen
  \bibfield  {author} {\bibinfo {author} {\bibfnamefont {L.}~\bibnamefont {Mandel}}\ and\ \bibinfo {author} {\bibfnamefont {E.}~\bibnamefont {Wolf}},\ }\href@noop {} {\emph {\bibinfo {title} {Optical coherence and quantum optics}}}\ (\bibinfo  {publisher} {Cambridge university press},\ \bibinfo {year} {1995})\BibitemShut {NoStop}%
\bibitem [{\citenamefont {Kimble}\ and\ \citenamefont {Mandel}(1976)}]{kimble1976theory}%
  \BibitemOpen
  \bibfield  {author} {\bibinfo {author} {\bibfnamefont {H.}~\bibnamefont {Kimble}}\ and\ \bibinfo {author} {\bibfnamefont {L.}~\bibnamefont {Mandel}},\ }\bibfield  {title} {\bibinfo {title} {Theory of resonance fluorescence},\ }\href {https://link.aps.org/doi/10.1103/PhysRevA.13.2123} {\bibfield  {journal} {\bibinfo  {journal} {Physical Review A}\ }\textbf {\bibinfo {volume} {13}},\ \bibinfo {pages} {2123} (\bibinfo {year} {1976})}\BibitemShut {NoStop}%
\bibitem [{\citenamefont {Kimble}\ \emph {et~al.}(1977)\citenamefont {Kimble}, \citenamefont {Dagenais},\ and\ \citenamefont {Mandel}}]{kimble1977photon}%
  \BibitemOpen
  \bibfield  {author} {\bibinfo {author} {\bibfnamefont {H.~J.}\ \bibnamefont {Kimble}}, \bibinfo {author} {\bibfnamefont {M.}~\bibnamefont {Dagenais}},\ and\ \bibinfo {author} {\bibfnamefont {L.}~\bibnamefont {Mandel}},\ }\bibfield  {title} {\bibinfo {title} {Photon antibunching in resonance fluorescence},\ }\href {https://link.aps.org/doi/10.1103/PhysRevLett.39.691} {\bibfield  {journal} {\bibinfo  {journal} {Physical Review Letters}\ }\textbf {\bibinfo {volume} {39}},\ \bibinfo {pages} {691} (\bibinfo {year} {1977})}\BibitemShut {NoStop}%
\bibitem [{\citenamefont {Itano}\ \emph {et~al.}(1988)\citenamefont {Itano}, \citenamefont {Bergquist},\ and\ \citenamefont {Wineland}}]{itano1988photon}%
  \BibitemOpen
  \bibfield  {author} {\bibinfo {author} {\bibfnamefont {W.~M.}\ \bibnamefont {Itano}}, \bibinfo {author} {\bibfnamefont {J.}~\bibnamefont {Bergquist}},\ and\ \bibinfo {author} {\bibfnamefont {D.}~\bibnamefont {Wineland}},\ }\bibfield  {title} {\bibinfo {title} {Photon antibunching and sub-poissonian statistics from quantum jumps in one and two atoms},\ }\href {https://journals.aps.org/pra/abstract/10.1103/PhysRevA.38.559} {\bibfield  {journal} {\bibinfo  {journal} {Physical Review A}\ }\textbf {\bibinfo {volume} {38}},\ \bibinfo {pages} {559} (\bibinfo {year} {1988})}\BibitemShut {NoStop}%
\bibitem [{\citenamefont {Brabec}\ and\ \citenamefont {Krausz}(2000)}]{brabec2000intense}%
  \BibitemOpen
  \bibfield  {author} {\bibinfo {author} {\bibfnamefont {T.}~\bibnamefont {Brabec}}\ and\ \bibinfo {author} {\bibfnamefont {F.}~\bibnamefont {Krausz}},\ }\bibfield  {title} {\bibinfo {title} {Intense few-cycle laser fields: Frontiers of nonlinear optics},\ }\href {https://link.aps.org/doi/10.1103/RevModPhys.72.545} {\bibfield  {journal} {\bibinfo  {journal} {Reviews of Modern Physics}\ }\textbf {\bibinfo {volume} {72}},\ \bibinfo {pages} {545} (\bibinfo {year} {2000})}\BibitemShut {NoStop}%
\bibitem [{\citenamefont {Krausz}\ and\ \citenamefont {Ivanov}(2009)}]{krausz2009attosecond}%
  \BibitemOpen
  \bibfield  {author} {\bibinfo {author} {\bibfnamefont {F.}~\bibnamefont {Krausz}}\ and\ \bibinfo {author} {\bibfnamefont {M.}~\bibnamefont {Ivanov}},\ }\bibfield  {title} {\bibinfo {title} {Attosecond physics},\ }\href {https://journals.aps.org/rmp/abstract/10.1103/RevModPhys.81.163} {\bibfield  {journal} {\bibinfo  {journal} {Reviews of Modern Physics}\ }\textbf {\bibinfo {volume} {81}},\ \bibinfo {pages} {163} (\bibinfo {year} {2009})}\BibitemShut {NoStop}%
\bibitem [{\citenamefont {Lewenstein}\ \emph {et~al.}(1994)\citenamefont {Lewenstein}, \citenamefont {Balcou}, \citenamefont {Ivanov}, \citenamefont {L’huillier},\ and\ \citenamefont {Corkum}}]{lewenstein1994theory}%
  \BibitemOpen
  \bibfield  {author} {\bibinfo {author} {\bibfnamefont {M.}~\bibnamefont {Lewenstein}}, \bibinfo {author} {\bibfnamefont {P.}~\bibnamefont {Balcou}}, \bibinfo {author} {\bibfnamefont {M.~Y.}\ \bibnamefont {Ivanov}}, \bibinfo {author} {\bibfnamefont {A.}~\bibnamefont {L’huillier}},\ and\ \bibinfo {author} {\bibfnamefont {P.~B.}\ \bibnamefont {Corkum}},\ }\bibfield  {title} {\bibinfo {title} {Theory of high-harmonic generation by low-frequency laser fields},\ }\href {https://link.aps.org/doi/10.1103/PhysRevA.49.2117} {\bibfield  {journal} {\bibinfo  {journal} {Physical Review A}\ }\textbf {\bibinfo {volume} {49}},\ \bibinfo {pages} {2117} (\bibinfo {year} {1994})}\BibitemShut {NoStop}%
\bibitem [{\citenamefont {Amini}\ \emph {et~al.}(2019)\citenamefont {Amini}, \citenamefont {Biegert}, \citenamefont {Calegari}, \citenamefont {Chac{\'o}n}, \citenamefont {Ciappina}, \citenamefont {Dauphin}, \citenamefont {Efimov}, \citenamefont {de~Morisson~Faria}, \citenamefont {Giergiel}, \citenamefont {Gniewek} \emph {et~al.}}]{amini2019symphony}%
  \BibitemOpen
  \bibfield  {author} {\bibinfo {author} {\bibfnamefont {K.}~\bibnamefont {Amini}}, \bibinfo {author} {\bibfnamefont {J.}~\bibnamefont {Biegert}}, \bibinfo {author} {\bibfnamefont {F.}~\bibnamefont {Calegari}}, \bibinfo {author} {\bibfnamefont {A.}~\bibnamefont {Chac{\'o}n}}, \bibinfo {author} {\bibfnamefont {M.~F.}\ \bibnamefont {Ciappina}}, \bibinfo {author} {\bibfnamefont {A.}~\bibnamefont {Dauphin}}, \bibinfo {author} {\bibfnamefont {D.~K.}\ \bibnamefont {Efimov}}, \bibinfo {author} {\bibfnamefont {C.~F.}\ \bibnamefont {de~Morisson~Faria}}, \bibinfo {author} {\bibfnamefont {K.}~\bibnamefont {Giergiel}}, \bibinfo {author} {\bibfnamefont {P.}~\bibnamefont {Gniewek}}, \emph {et~al.},\ }\bibfield  {title} {\bibinfo {title} {Symphony on strong field approximation},\ }\href {https://dx.doi.org/10.1088/1361-6633/ab2bb1} {\bibfield  {journal} {\bibinfo  {journal} {Reports on Progress in Physics}\ }\textbf {\bibinfo {volume} {82}},\ \bibinfo {pages} {116001} (\bibinfo {year} {2019})}\BibitemShut {NoStop}%
\bibitem [{\citenamefont {Lewenstein}\ \emph {et~al.}(2021)\citenamefont {Lewenstein}, \citenamefont {Ciappina}, \citenamefont {Pisanty}, \citenamefont {Rivera-Dean}, \citenamefont {Stammer}, \citenamefont {Lamprou},\ and\ \citenamefont {Tzallas}}]{lewenstein2021generation}%
  \BibitemOpen
  \bibfield  {author} {\bibinfo {author} {\bibfnamefont {M.}~\bibnamefont {Lewenstein}}, \bibinfo {author} {\bibfnamefont {M.~F.}\ \bibnamefont {Ciappina}}, \bibinfo {author} {\bibfnamefont {E.}~\bibnamefont {Pisanty}}, \bibinfo {author} {\bibfnamefont {J.}~\bibnamefont {Rivera-Dean}}, \bibinfo {author} {\bibfnamefont {P.}~\bibnamefont {Stammer}}, \bibinfo {author} {\bibfnamefont {T.}~\bibnamefont {Lamprou}},\ and\ \bibinfo {author} {\bibfnamefont {P.}~\bibnamefont {Tzallas}},\ }\bibfield  {title} {\bibinfo {title} {Generation of optical schr{\"o}dinger cat states in intense laser--matter interactions},\ }\href {https://www.nature.com/articles/s41567-021-01317-w} {\bibfield  {journal} {\bibinfo  {journal} {Nature Physics}\ }\textbf {\bibinfo {volume} {17}},\ \bibinfo {pages} {1104} (\bibinfo {year} {2021})}\BibitemShut {NoStop}%
\bibitem [{\citenamefont {Gorlach}\ \emph {et~al.}(2020)\citenamefont {Gorlach}, \citenamefont {Neufeld}, \citenamefont {Rivera}, \citenamefont {Cohen},\ and\ \citenamefont {Kaminer}}]{gorlach2020quantum}%
  \BibitemOpen
  \bibfield  {author} {\bibinfo {author} {\bibfnamefont {A.}~\bibnamefont {Gorlach}}, \bibinfo {author} {\bibfnamefont {O.}~\bibnamefont {Neufeld}}, \bibinfo {author} {\bibfnamefont {N.}~\bibnamefont {Rivera}}, \bibinfo {author} {\bibfnamefont {O.}~\bibnamefont {Cohen}},\ and\ \bibinfo {author} {\bibfnamefont {I.}~\bibnamefont {Kaminer}},\ }\bibfield  {title} {\bibinfo {title} {The quantum-optical nature of high harmonic generation},\ }\href {https://www.nature.com/articles/s41467-020-18218-w} {\bibfield  {journal} {\bibinfo  {journal} {Nat. Commun.}\ }\textbf {\bibinfo {volume} {11}},\ \bibinfo {pages} {4598} (\bibinfo {year} {2020})}\BibitemShut {NoStop}%
\bibitem [{\citenamefont {Stammer}\ \emph {et~al.}(2025{\natexlab{a}})\citenamefont {Stammer}, \citenamefont {Rivera-Dean}, \citenamefont {Tzallas}, \citenamefont {Ciappina},\ and\ \citenamefont {Lewenstein}}]{stammer2025colloquium}%
  \BibitemOpen
  \bibfield  {author} {\bibinfo {author} {\bibfnamefont {P.}~\bibnamefont {Stammer}}, \bibinfo {author} {\bibfnamefont {J.}~\bibnamefont {Rivera-Dean}}, \bibinfo {author} {\bibfnamefont {P.}~\bibnamefont {Tzallas}}, \bibinfo {author} {\bibfnamefont {M.~F.}\ \bibnamefont {Ciappina}},\ and\ \bibinfo {author} {\bibfnamefont {M.}~\bibnamefont {Lewenstein}},\ }\bibfield  {title} {\bibinfo {title} {Colloquium: Quantum optics of intense light--matter interaction},\ }\href {https://arxiv.org/abs/2510.19045} {\bibfield  {journal} {\bibinfo  {journal} {arXiv:2510.19045}\ } (\bibinfo {year} {2025}{\natexlab{a}})}\BibitemShut {NoStop}%
\bibitem [{\citenamefont {Cruz-Rodriguez}\ \emph {et~al.}(2024)\citenamefont {Cruz-Rodriguez}, \citenamefont {Dey}, \citenamefont {Freibert},\ and\ \citenamefont {Stammer}}]{cruz2024quantum}%
  \BibitemOpen
  \bibfield  {author} {\bibinfo {author} {\bibfnamefont {L.}~\bibnamefont {Cruz-Rodriguez}}, \bibinfo {author} {\bibfnamefont {D.}~\bibnamefont {Dey}}, \bibinfo {author} {\bibfnamefont {A.}~\bibnamefont {Freibert}},\ and\ \bibinfo {author} {\bibfnamefont {P.}~\bibnamefont {Stammer}},\ }\bibfield  {title} {\bibinfo {title} {Quantum phenomena in attosecond science},\ }\href {https://doi.org/10.1038/s42254-024-00769-2} {\bibfield  {journal} {\bibinfo  {journal} {Nat. Rev. Phys.}\ }\textbf {\bibinfo {volume} {6}},\ \bibinfo {pages} {691} (\bibinfo {year} {2024})}\BibitemShut {NoStop}%
\bibitem [{\citenamefont {Stammer}\ \emph {et~al.}(2022)\citenamefont {Stammer}, \citenamefont {Rivera-Dean}, \citenamefont {Lamprou}, \citenamefont {Pisanty}, \citenamefont {Ciappina}, \citenamefont {Tzallas},\ and\ \citenamefont {Lewenstein}}]{stammer2022high}%
  \BibitemOpen
  \bibfield  {author} {\bibinfo {author} {\bibfnamefont {P.}~\bibnamefont {Stammer}}, \bibinfo {author} {\bibfnamefont {J.}~\bibnamefont {Rivera-Dean}}, \bibinfo {author} {\bibfnamefont {T.}~\bibnamefont {Lamprou}}, \bibinfo {author} {\bibfnamefont {E.}~\bibnamefont {Pisanty}}, \bibinfo {author} {\bibfnamefont {M.~F.}\ \bibnamefont {Ciappina}}, \bibinfo {author} {\bibfnamefont {P.}~\bibnamefont {Tzallas}},\ and\ \bibinfo {author} {\bibfnamefont {M.}~\bibnamefont {Lewenstein}},\ }\bibfield  {title} {\bibinfo {title} {High photon number entangled states and coherent state superposition from the extreme ultraviolet to the far infrared},\ }\href {https://link.aps.org/doi/10.1103/PhysRevLett.128.123603} {\bibfield  {journal} {\bibinfo  {journal} {Physical Review Letters}\ }\textbf {\bibinfo {volume} {128}},\ \bibinfo {pages} {123603} (\bibinfo {year} {2022})}\BibitemShut {NoStop}%
\bibitem [{\citenamefont {Stammer}(2022)}]{stammer2022theory}%
  \BibitemOpen
  \bibfield  {author} {\bibinfo {author} {\bibfnamefont {P.}~\bibnamefont {Stammer}},\ }\bibfield  {title} {\bibinfo {title} {Theory of entanglement and measurement in high-order harmonic generation},\ }\href {https://link.aps.org/doi/10.1103/PhysRevA.106.L050402} {\bibfield  {journal} {\bibinfo  {journal} {Physical Review A}\ }\textbf {\bibinfo {volume} {106}},\ \bibinfo {pages} {L050402} (\bibinfo {year} {2022})}\BibitemShut {NoStop}%
\bibitem [{\citenamefont {Yi}\ \emph {et~al.}(2025)\citenamefont {Yi}, \citenamefont {Klimkin}, \citenamefont {Brown}, \citenamefont {Smirnova}, \citenamefont {Patchkovskii}, \citenamefont {Babushkin},\ and\ \citenamefont {Ivanov}}]{yi2024generation}%
  \BibitemOpen
  \bibfield  {author} {\bibinfo {author} {\bibfnamefont {S.}~\bibnamefont {Yi}}, \bibinfo {author} {\bibfnamefont {N.~D.}\ \bibnamefont {Klimkin}}, \bibinfo {author} {\bibfnamefont {G.~G.}\ \bibnamefont {Brown}}, \bibinfo {author} {\bibfnamefont {O.}~\bibnamefont {Smirnova}}, \bibinfo {author} {\bibfnamefont {S.}~\bibnamefont {Patchkovskii}}, \bibinfo {author} {\bibfnamefont {I.}~\bibnamefont {Babushkin}},\ and\ \bibinfo {author} {\bibfnamefont {M.}~\bibnamefont {Ivanov}},\ }\bibfield  {title} {\bibinfo {title} {Generation of massively entangled bright states of light during harmonic generation in resonant media},\ }\href {https://link.aps.org/doi/10.1103/PhysRevX.15.011023} {\bibfield  {journal} {\bibinfo  {journal} {Physical Review X}\ }\textbf {\bibinfo {volume} {15}},\ \bibinfo {pages} {011023} (\bibinfo {year} {2025})}\BibitemShut {NoStop}%
\bibitem [{\citenamefont {Stammer}\ \emph {et~al.}(2024)\citenamefont {Stammer}, \citenamefont {Rivera-Dean}, \citenamefont {Maxwell}, \citenamefont {Lamprou}, \citenamefont {Arg{\"u}ello-Luengo}, \citenamefont {Tzallas}, \citenamefont {Ciappina},\ and\ \citenamefont {Lewenstein}}]{stammer2024entanglement}%
  \BibitemOpen
  \bibfield  {author} {\bibinfo {author} {\bibfnamefont {P.}~\bibnamefont {Stammer}}, \bibinfo {author} {\bibfnamefont {J.}~\bibnamefont {Rivera-Dean}}, \bibinfo {author} {\bibfnamefont {A.~S.}\ \bibnamefont {Maxwell}}, \bibinfo {author} {\bibfnamefont {T.}~\bibnamefont {Lamprou}}, \bibinfo {author} {\bibfnamefont {J.}~\bibnamefont {Arg{\"u}ello-Luengo}}, \bibinfo {author} {\bibfnamefont {P.}~\bibnamefont {Tzallas}}, \bibinfo {author} {\bibfnamefont {M.~F.}\ \bibnamefont {Ciappina}},\ and\ \bibinfo {author} {\bibfnamefont {M.}~\bibnamefont {Lewenstein}},\ }\bibfield  {title} {\bibinfo {title} {Entanglement and squeezing of the optical field modes in high harmonic generation},\ }\href {https://journals.aps.org/prl/abstract/10.1103/PhysRevLett.132.143603} {\bibfield  {journal} {\bibinfo  {journal} {Phys. Rev. Lett.}\ }\textbf {\bibinfo {volume} {132}},\ \bibinfo {pages} {143603} (\bibinfo {year} {2024})}\BibitemShut {NoStop}%
\bibitem [{\citenamefont {Lange}\ \emph {et~al.}(2025)\citenamefont {Lange}, \citenamefont {Hansen},\ and\ \citenamefont {Madsen}}]{lange_excitonic_2025}%
  \BibitemOpen
  \bibfield  {author} {\bibinfo {author} {\bibfnamefont {C.~S.}\ \bibnamefont {Lange}}, \bibinfo {author} {\bibfnamefont {T.}~\bibnamefont {Hansen}},\ and\ \bibinfo {author} {\bibfnamefont {L.~B.}\ \bibnamefont {Madsen}},\ }\bibfield  {title} {\bibinfo {title} {Excitonic {Enhancement} of {Squeezed} {Light} in {Quantum}-{Optical} {High}-{Harmonic} {Generation} from a {Mott} {Insulator}},\ }\href {https://doi.org/10.1103/wyk5-k8tk} {\bibfield  {journal} {\bibinfo  {journal} {Phys. Rev. Lett.}\ }\textbf {\bibinfo {volume} {135}},\ \bibinfo {pages} {043603} (\bibinfo {year} {2025})}\BibitemShut {NoStop}%
\bibitem [{\citenamefont {Tzur}\ \emph {et~al.}(2024)\citenamefont {Tzur}, \citenamefont {Birk}, \citenamefont {Gorlach}, \citenamefont {Kaminer}, \citenamefont {Kr{\"u}ger},\ and\ \citenamefont {Cohen}}]{tzur2024generation}%
  \BibitemOpen
  \bibfield  {author} {\bibinfo {author} {\bibfnamefont {M.~E.}\ \bibnamefont {Tzur}}, \bibinfo {author} {\bibfnamefont {M.}~\bibnamefont {Birk}}, \bibinfo {author} {\bibfnamefont {A.}~\bibnamefont {Gorlach}}, \bibinfo {author} {\bibfnamefont {I.}~\bibnamefont {Kaminer}}, \bibinfo {author} {\bibfnamefont {M.}~\bibnamefont {Kr{\"u}ger}},\ and\ \bibinfo {author} {\bibfnamefont {O.}~\bibnamefont {Cohen}},\ }\bibfield  {title} {\bibinfo {title} {Generation of squeezed high-order harmonics},\ }\href {https://link.aps.org/doi/10.1103/PhysRevResearch.6.033079} {\bibfield  {journal} {\bibinfo  {journal} {Physical Review Research}\ }\textbf {\bibinfo {volume} {6}},\ \bibinfo {pages} {033079} (\bibinfo {year} {2024})}\BibitemShut {NoStop}%
\bibitem [{\citenamefont {Rivera-Dean}\ \emph {et~al.}(2022)\citenamefont {Rivera-Dean}, \citenamefont {Lamprou}, \citenamefont {Pisanty}, \citenamefont {Stammer}, \citenamefont {Ord{\'o}{\~n}ez}, \citenamefont {Maxwell}, \citenamefont {Ciappina}, \citenamefont {Lewenstein},\ and\ \citenamefont {Tzallas}}]{rivera2022strong}%
  \BibitemOpen
  \bibfield  {author} {\bibinfo {author} {\bibfnamefont {J.}~\bibnamefont {Rivera-Dean}}, \bibinfo {author} {\bibfnamefont {T.}~\bibnamefont {Lamprou}}, \bibinfo {author} {\bibfnamefont {E.}~\bibnamefont {Pisanty}}, \bibinfo {author} {\bibfnamefont {P.}~\bibnamefont {Stammer}}, \bibinfo {author} {\bibfnamefont {A.~F.}\ \bibnamefont {Ord{\'o}{\~n}ez}}, \bibinfo {author} {\bibfnamefont {A.~S.}\ \bibnamefont {Maxwell}}, \bibinfo {author} {\bibfnamefont {M.~F.}\ \bibnamefont {Ciappina}}, \bibinfo {author} {\bibfnamefont {M.}~\bibnamefont {Lewenstein}},\ and\ \bibinfo {author} {\bibfnamefont {P.}~\bibnamefont {Tzallas}},\ }\bibfield  {title} {\bibinfo {title} {Strong laser fields and their power to generate controllable high-photon-number coherent-state superpositions},\ }\href {https://link.aps.org/doi/10.1103/PhysRevA.105.033714} {\bibfield  {journal} {\bibinfo  {journal} {Physical Review A}\ }\textbf {\bibinfo {volume} {105}},\ \bibinfo {pages} {033714} (\bibinfo {year} {2022})}\BibitemShut {NoStop}%
\bibitem [{\citenamefont {Theidel}\ \emph {et~al.}(2024)\citenamefont {Theidel}, \citenamefont {Cotte}, \citenamefont {Sondenheimer}, \citenamefont {Shiriaeva}, \citenamefont {Froidevaux}, \citenamefont {Severin}, \citenamefont {Merdji-Larue}, \citenamefont {Mosel}, \citenamefont {Fr{\"o}hlich}, \citenamefont {Weber} \emph {et~al.}}]{theidel2024evidence}%
  \BibitemOpen
  \bibfield  {author} {\bibinfo {author} {\bibfnamefont {D.}~\bibnamefont {Theidel}}, \bibinfo {author} {\bibfnamefont {V.}~\bibnamefont {Cotte}}, \bibinfo {author} {\bibfnamefont {R.}~\bibnamefont {Sondenheimer}}, \bibinfo {author} {\bibfnamefont {V.}~\bibnamefont {Shiriaeva}}, \bibinfo {author} {\bibfnamefont {M.}~\bibnamefont {Froidevaux}}, \bibinfo {author} {\bibfnamefont {V.}~\bibnamefont {Severin}}, \bibinfo {author} {\bibfnamefont {A.}~\bibnamefont {Merdji-Larue}}, \bibinfo {author} {\bibfnamefont {P.}~\bibnamefont {Mosel}}, \bibinfo {author} {\bibfnamefont {S.}~\bibnamefont {Fr{\"o}hlich}}, \bibinfo {author} {\bibfnamefont {K.-A.}\ \bibnamefont {Weber}}, \emph {et~al.},\ }\bibfield  {title} {\bibinfo {title} {Evidence of the quantum optical nature of high-harmonic generation},\ }\href {https://link.aps.org/doi/10.1103/PRXQuantum.5.040319} {\bibfield  {journal} {\bibinfo  {journal} {PRX Quantum}\ }\textbf {\bibinfo {volume} {5}},\ \bibinfo {pages} {040319} (\bibinfo {year} {2024})}\BibitemShut
  {NoStop}%
\bibitem [{\citenamefont {Lemieux}\ \emph {et~al.}(2025)\citenamefont {Lemieux}, \citenamefont {Jalil}, \citenamefont {Purschke}, \citenamefont {Boroumand}, \citenamefont {Hammond}, \citenamefont {Villeneuve}, \citenamefont {Naumov}, \citenamefont {Brabec},\ and\ \citenamefont {Vampa}}]{lemieux2024photon}%
  \BibitemOpen
  \bibfield  {author} {\bibinfo {author} {\bibfnamefont {S.}~\bibnamefont {Lemieux}}, \bibinfo {author} {\bibfnamefont {S.~A.}\ \bibnamefont {Jalil}}, \bibinfo {author} {\bibfnamefont {D.~N.}\ \bibnamefont {Purschke}}, \bibinfo {author} {\bibfnamefont {N.}~\bibnamefont {Boroumand}}, \bibinfo {author} {\bibfnamefont {T.}~\bibnamefont {Hammond}}, \bibinfo {author} {\bibfnamefont {D.}~\bibnamefont {Villeneuve}}, \bibinfo {author} {\bibfnamefont {A.}~\bibnamefont {Naumov}}, \bibinfo {author} {\bibfnamefont {T.}~\bibnamefont {Brabec}},\ and\ \bibinfo {author} {\bibfnamefont {G.}~\bibnamefont {Vampa}},\ }\bibfield  {title} {\bibinfo {title} {Photon bunching in high-harmonic emission controlled by quantum light},\ }\href {https://www.nature.com/articles/s41566-025-01673-6} {\bibfield  {journal} {\bibinfo  {journal} {Nature Photonics}\ ,\ \bibinfo {pages} {1}} (\bibinfo {year} {2025})}\BibitemShut {NoStop}%
\bibitem [{\citenamefont {Zou}\ and\ \citenamefont {Mandel}(1990)}]{zou1990photon}%
  \BibitemOpen
  \bibfield  {author} {\bibinfo {author} {\bibfnamefont {X.~T.}\ \bibnamefont {Zou}}\ and\ \bibinfo {author} {\bibfnamefont {L.}~\bibnamefont {Mandel}},\ }\bibfield  {title} {\bibinfo {title} {Photon-antibunching and sub-poissonian photon statistics},\ }\href {https://doi.org/10.1103/PhysRevA.41.475} {\bibfield  {journal} {\bibinfo  {journal} {Phys. Rev. A}\ }\textbf {\bibinfo {volume} {41}},\ \bibinfo {pages} {475} (\bibinfo {year} {1990})}\BibitemShut {NoStop}%
\bibitem [{\citenamefont {Stammer}\ \emph {et~al.}(2025{\natexlab{b}})\citenamefont {Stammer}, \citenamefont {Rivera-Dean},\ and\ \citenamefont {Lewenstein}}]{stammer2025theory}%
  \BibitemOpen
  \bibfield  {author} {\bibinfo {author} {\bibfnamefont {P.}~\bibnamefont {Stammer}}, \bibinfo {author} {\bibfnamefont {J.}~\bibnamefont {Rivera-Dean}},\ and\ \bibinfo {author} {\bibfnamefont {M.}~\bibnamefont {Lewenstein}},\ }\bibfield  {title} {\bibinfo {title} {Theory of quantum optics and optical coherence in high harmonic generation},\ }\href {https://arxiv.org/abs/2504.13287} {\bibfield  {journal} {\bibinfo  {journal} {arXiv:2504.13287}\ } (\bibinfo {year} {2025}{\natexlab{b}})}\BibitemShut {NoStop}%
\bibitem [{\citenamefont {Sundaram}\ and\ \citenamefont {Milonni}(1990)}]{sundaram1990high}%
  \BibitemOpen
  \bibfield  {author} {\bibinfo {author} {\bibfnamefont {B.}~\bibnamefont {Sundaram}}\ and\ \bibinfo {author} {\bibfnamefont {P.~W.}\ \bibnamefont {Milonni}},\ }\bibfield  {title} {\bibinfo {title} {High-order harmonic generation: simplified model and relevance of single-atom theories to experiment},\ }\href {https://link.aps.org/doi/10.1103/PhysRevA.41.6571} {\bibfield  {journal} {\bibinfo  {journal} {Physical Review A}\ }\textbf {\bibinfo {volume} {41}},\ \bibinfo {pages} {6571} (\bibinfo {year} {1990})}\BibitemShut {NoStop}%
\bibitem [{\citenamefont {Diestler}(2008)}]{diestler2008harmonic}%
  \BibitemOpen
  \bibfield  {author} {\bibinfo {author} {\bibfnamefont {D.}~\bibnamefont {Diestler}},\ }\bibfield  {title} {\bibinfo {title} {Harmonic generation: quantum-electrodynamical theory of the harmonic photon-number spectrum},\ }\href {https://link.aps.org/doi/10.1103/PhysRevA.78.033814} {\bibfield  {journal} {\bibinfo  {journal} {Physical Review A—Atomic, Molecular, and Optical Physics}\ }\textbf {\bibinfo {volume} {78}},\ \bibinfo {pages} {033814} (\bibinfo {year} {2008})}\BibitemShut {NoStop}%
\bibitem [{\citenamefont {Stammer}\ \emph {et~al.}(2023)\citenamefont {Stammer}, \citenamefont {Rivera-Dean}, \citenamefont {Maxwell}, \citenamefont {Lamprou}, \citenamefont {Ord{\'o}{\~n}ez}, \citenamefont {Ciappina}, \citenamefont {Tzallas},\ and\ \citenamefont {Lewenstein}}]{stammer2023quantum}%
  \BibitemOpen
  \bibfield  {author} {\bibinfo {author} {\bibfnamefont {P.}~\bibnamefont {Stammer}}, \bibinfo {author} {\bibfnamefont {J.}~\bibnamefont {Rivera-Dean}}, \bibinfo {author} {\bibfnamefont {A.}~\bibnamefont {Maxwell}}, \bibinfo {author} {\bibfnamefont {T.}~\bibnamefont {Lamprou}}, \bibinfo {author} {\bibfnamefont {A.}~\bibnamefont {Ord{\'o}{\~n}ez}}, \bibinfo {author} {\bibfnamefont {M.~F.}\ \bibnamefont {Ciappina}}, \bibinfo {author} {\bibfnamefont {P.}~\bibnamefont {Tzallas}},\ and\ \bibinfo {author} {\bibfnamefont {M.}~\bibnamefont {Lewenstein}},\ }\bibfield  {title} {\bibinfo {title} {Quantum electrodynamics of intense laser-matter interactions: a tool for quantum state engineering},\ }\href {https://link.aps.org/doi/10.1103/PRXQuantum.4.010201} {\bibfield  {journal} {\bibinfo  {journal} {PRX Quantum}\ }\textbf {\bibinfo {volume} {4}},\ \bibinfo {pages} {010201} (\bibinfo {year} {2023})}\BibitemShut {NoStop}%
\bibitem [{\citenamefont {Wiener}(1930)}]{wiener1930generalized}%
  \BibitemOpen
  \bibfield  {author} {\bibinfo {author} {\bibfnamefont {N.}~\bibnamefont {Wiener}},\ }\bibfield  {title} {\bibinfo {title} {Generalized harmonic analysis},\ }\href {https://link.springer.com/article/10.1007/BF02546511} {\bibfield  {journal} {\bibinfo  {journal} {Acta mathematica}\ }\textbf {\bibinfo {volume} {55}},\ \bibinfo {pages} {117} (\bibinfo {year} {1930})}\BibitemShut {NoStop}%
\bibitem [{\citenamefont {Khintchine}(1934)}]{khintchine1934korrelationstheorie}%
  \BibitemOpen
  \bibfield  {author} {\bibinfo {author} {\bibfnamefont {A.}~\bibnamefont {Khintchine}},\ }\bibfield  {title} {\bibinfo {title} {Korrelationstheorie der station{\"a}ren stochastischen prozesse},\ }\href {https://link.springer.com/article/10.1007/bf01449156} {\bibfield  {journal} {\bibinfo  {journal} {Mathematische Annalen}\ }\textbf {\bibinfo {volume} {109}},\ \bibinfo {pages} {604} (\bibinfo {year} {1934})}\BibitemShut {NoStop}%
\bibitem [{\citenamefont {Stammer}(2025{\natexlab{a}})}]{stammer2025quantum}%
  \BibitemOpen
  \bibfield  {author} {\bibinfo {author} {\bibfnamefont {P.}~\bibnamefont {Stammer}},\ }\bibfield  {title} {\bibinfo {title} {Quantum stochastic analysis of non-linear driven light emission},\ }\href {https://arxiv.org/abs/2508.09049} {\bibfield  {journal} {\bibinfo  {journal} {arXiv:2508.09049}\ } (\bibinfo {year} {2025}{\natexlab{a}})}\BibitemShut {NoStop}%
\bibitem [{\citenamefont {Brown}\ and\ \citenamefont {Twiss}(1956)}]{brown1956correlation}%
  \BibitemOpen
  \bibfield  {author} {\bibinfo {author} {\bibfnamefont {R.~H.}\ \bibnamefont {Brown}}\ and\ \bibinfo {author} {\bibfnamefont {R.~Q.}\ \bibnamefont {Twiss}},\ }\bibfield  {title} {\bibinfo {title} {Correlation between photons in two coherent beams of light},\ }\href {https://www.nature.com/articles/177027a0} {\bibfield  {journal} {\bibinfo  {journal} {Nature}\ }\textbf {\bibinfo {volume} {177}},\ \bibinfo {pages} {27} (\bibinfo {year} {1956})}\BibitemShut {NoStop}%
\bibitem [{\citenamefont {Mandel}\ and\ \citenamefont {Wolf}(1966)}]{mandel1966photon}%
  \BibitemOpen
  \bibfield  {author} {\bibinfo {author} {\bibfnamefont {L.}~\bibnamefont {Mandel}}\ and\ \bibinfo {author} {\bibfnamefont {E.}~\bibnamefont {Wolf}},\ }\bibfield  {title} {\bibinfo {title} {Photon statistics and classical fields},\ }\href {https://journals.aps.org/pr/abstract/10.1103/PhysRev.149.1033} {\bibfield  {journal} {\bibinfo  {journal} {Physical Review}\ }\textbf {\bibinfo {volume} {149}},\ \bibinfo {pages} {1033} (\bibinfo {year} {1966})}\BibitemShut {NoStop}%
\bibitem [{\citenamefont {Corkum}(1993)}]{corkum1993plasma}%
  \BibitemOpen
  \bibfield  {author} {\bibinfo {author} {\bibfnamefont {P.~B.}\ \bibnamefont {Corkum}},\ }\bibfield  {title} {\bibinfo {title} {Plasma perspective on strong field multiphoton ionization},\ }\href {https://link.aps.org/doi/10.1103/PhysRevLett.71.1994} {\bibfield  {journal} {\bibinfo  {journal} {Physical review letters}\ }\textbf {\bibinfo {volume} {71}},\ \bibinfo {pages} {1994} (\bibinfo {year} {1993})}\BibitemShut {NoStop}%
\bibitem [{\citenamefont {Salières}\ \emph {et~al.}(2001)\citenamefont {Salières}, \citenamefont {Carré}, \citenamefont {Le~Déroff}, \citenamefont {Grasbon}, \citenamefont {Paulus}, \citenamefont {Walther}, \citenamefont {Kopold}, \citenamefont {Becker}, \citenamefont {Milošević}, \citenamefont {Sanpera},\ and\ \citenamefont {Lewenstein}}]{salieres_feynmans_2001}%
  \BibitemOpen
  \bibfield  {author} {\bibinfo {author} {\bibfnamefont {P.}~\bibnamefont {Salières}}, \bibinfo {author} {\bibfnamefont {B.}~\bibnamefont {Carré}}, \bibinfo {author} {\bibfnamefont {L.}~\bibnamefont {Le~Déroff}}, \bibinfo {author} {\bibfnamefont {F.}~\bibnamefont {Grasbon}}, \bibinfo {author} {\bibfnamefont {G.~G.}\ \bibnamefont {Paulus}}, \bibinfo {author} {\bibfnamefont {H.}~\bibnamefont {Walther}}, \bibinfo {author} {\bibfnamefont {R.}~\bibnamefont {Kopold}}, \bibinfo {author} {\bibfnamefont {W.}~\bibnamefont {Becker}}, \bibinfo {author} {\bibfnamefont {D.~B.}\ \bibnamefont {Milošević}}, \bibinfo {author} {\bibfnamefont {A.}~\bibnamefont {Sanpera}},\ and\ \bibinfo {author} {\bibfnamefont {M.}~\bibnamefont {Lewenstein}},\ }\bibfield  {title} {\bibinfo {title} {Feynman's {Path}-{Integral} {Approach} for {Intense}-{Laser}-{Atom} {Interactions}},\ }\href {https://doi.org/10.1126/science.108836} {\bibfield  {journal} {\bibinfo  {journal} {Science}\ }\textbf {\bibinfo {volume} {292}},\ \bibinfo {pages}
  {902} (\bibinfo {year} {2001})}\BibitemShut {NoStop}%
\bibitem [{\citenamefont {Carmichael}(2013)}]{carmichael2013statistical}%
  \BibitemOpen
  \bibfield  {author} {\bibinfo {author} {\bibfnamefont {H.~J.}\ \bibnamefont {Carmichael}},\ }\href@noop {} {\emph {\bibinfo {title} {{Statistical Methods in Quantum Optics 1: Master Equations and Fokker-Planck Equations}}}}\ (\bibinfo  {publisher} {Springer-Verlag},\ \bibinfo {address} {Heidelberg},\ \bibinfo {year} {2013})\BibitemShut {NoStop}%
\bibitem [{\citenamefont {Boitier}\ \emph {et~al.}(2009)\citenamefont {Boitier}, \citenamefont {Godard}, \citenamefont {Rosencher},\ and\ \citenamefont {Fabre}}]{boitier2009measuring}%
  \BibitemOpen
  \bibfield  {author} {\bibinfo {author} {\bibfnamefont {F.}~\bibnamefont {Boitier}}, \bibinfo {author} {\bibfnamefont {A.}~\bibnamefont {Godard}}, \bibinfo {author} {\bibfnamefont {E.}~\bibnamefont {Rosencher}},\ and\ \bibinfo {author} {\bibfnamefont {C.}~\bibnamefont {Fabre}},\ }\bibfield  {title} {\bibinfo {title} {Measuring photon bunching at ultrashort timescale by two-photon absorption in semiconductors},\ }\href {https://www.nature.com/articles/nphys1218} {\bibfield  {journal} {\bibinfo  {journal} {Nature Physics}\ }\textbf {\bibinfo {volume} {5}},\ \bibinfo {pages} {267} (\bibinfo {year} {2009})}\BibitemShut {NoStop}%
\bibitem [{\citenamefont {Rivera-Dean}\ \emph {et~al.}(2026)\citenamefont {Rivera-Dean}, \citenamefont {Petrovic}, \citenamefont {Lewenstein},\ and\ \citenamefont {Stammer}}]{rivera2026attosecond}%
  \BibitemOpen
  \bibfield  {author} {\bibinfo {author} {\bibfnamefont {J.}~\bibnamefont {Rivera-Dean}}, \bibinfo {author} {\bibfnamefont {L.}~\bibnamefont {Petrovic}}, \bibinfo {author} {\bibfnamefont {M.}~\bibnamefont {Lewenstein}},\ and\ \bibinfo {author} {\bibfnamefont {P.}~\bibnamefont {Stammer}},\ }\bibfield  {title} {\bibinfo {title} {Attosecond quantum optical interferometry},\ }\href {https://iopscience.iop.org/article/10.1088/1361-6633/ae5847/meta} {\bibfield  {journal} {\bibinfo  {journal} {Reports on Progress in Physics}\ }\textbf {\bibinfo {volume} {89}},\ \bibinfo {pages} {047901} (\bibinfo {year} {2026})}\BibitemShut {NoStop}%
\bibitem [{\citenamefont {Lamprou}\ \emph {et~al.}(2025)\citenamefont {Lamprou}, \citenamefont {Rivera-Dean}, \citenamefont {Stammer}, \citenamefont {Lewenstein},\ and\ \citenamefont {Tzallas}}]{lamprou2025nonlinear}%
  \BibitemOpen
  \bibfield  {author} {\bibinfo {author} {\bibfnamefont {T.}~\bibnamefont {Lamprou}}, \bibinfo {author} {\bibfnamefont {J.}~\bibnamefont {Rivera-Dean}}, \bibinfo {author} {\bibfnamefont {P.}~\bibnamefont {Stammer}}, \bibinfo {author} {\bibfnamefont {M.}~\bibnamefont {Lewenstein}},\ and\ \bibinfo {author} {\bibfnamefont {P.}~\bibnamefont {Tzallas}},\ }\bibfield  {title} {\bibinfo {title} {Nonlinear optics using intense optical coherent state superpositions},\ }\href {https://journals.aps.org/prl/abstract/10.1103/PhysRevLett.134.013601} {\bibfield  {journal} {\bibinfo  {journal} {Phys. Rev. Lett.}\ }\textbf {\bibinfo {volume} {134}},\ \bibinfo {pages} {013601} (\bibinfo {year} {2025})}\BibitemShut {NoStop}%
\bibitem [{\citenamefont {Tzur}\ \emph {et~al.}(2025)\citenamefont {Tzur}, \citenamefont {Mor}, \citenamefont {Yaffe}, \citenamefont {Birk}, \citenamefont {Rasputnyi}, \citenamefont {Kneller}, \citenamefont {Nisim}, \citenamefont {Kaminer}, \citenamefont {Kr{\"u}ger}, \citenamefont {Dudovich} \emph {et~al.}}]{tzur2025measuring}%
  \BibitemOpen
  \bibfield  {author} {\bibinfo {author} {\bibfnamefont {M.~E.}\ \bibnamefont {Tzur}}, \bibinfo {author} {\bibfnamefont {C.}~\bibnamefont {Mor}}, \bibinfo {author} {\bibfnamefont {N.}~\bibnamefont {Yaffe}}, \bibinfo {author} {\bibfnamefont {M.}~\bibnamefont {Birk}}, \bibinfo {author} {\bibfnamefont {A.}~\bibnamefont {Rasputnyi}}, \bibinfo {author} {\bibfnamefont {O.}~\bibnamefont {Kneller}}, \bibinfo {author} {\bibfnamefont {I.}~\bibnamefont {Nisim}}, \bibinfo {author} {\bibfnamefont {I.}~\bibnamefont {Kaminer}}, \bibinfo {author} {\bibfnamefont {M.}~\bibnamefont {Kr{\"u}ger}}, \bibinfo {author} {\bibfnamefont {N.}~\bibnamefont {Dudovich}}, \emph {et~al.},\ }\bibfield  {title} {\bibinfo {title} {Measuring and controlling the birth of quantum attosecond pulses},\ }\href {https://arxiv.org/abs/2502.09427} {\bibfield  {journal} {\bibinfo  {journal} {arXiv:2502.09427}\ } (\bibinfo {year} {2025})}\BibitemShut {NoStop}%
\bibitem [{\citenamefont {Stammer}\ \emph {et~al.}(2025{\natexlab{c}})\citenamefont {Stammer}, \citenamefont {Rivera-Dean}, \citenamefont {Ciappina},\ and\ \citenamefont {Lewenstein}}]{stammer2025weak}%
  \BibitemOpen
  \bibfield  {author} {\bibinfo {author} {\bibfnamefont {P.}~\bibnamefont {Stammer}}, \bibinfo {author} {\bibfnamefont {J.}~\bibnamefont {Rivera-Dean}}, \bibinfo {author} {\bibfnamefont {M.~F.}\ \bibnamefont {Ciappina}},\ and\ \bibinfo {author} {\bibfnamefont {M.}~\bibnamefont {Lewenstein}},\ }\bibfield  {title} {\bibinfo {title} {Weak measurement in strong laser field physics},\ }\href {https://arxiv.org/abs/2508.09048} {\bibfield  {journal} {\bibinfo  {journal} {arXiv:2508.09048}\ } (\bibinfo {year} {2025}{\natexlab{c}})}\BibitemShut {NoStop}%
\bibitem [{\citenamefont {Pizzi}\ \emph {et~al.}(2023)\citenamefont {Pizzi}, \citenamefont {Gorlach}, \citenamefont {Rivera}, \citenamefont {Nunnenkamp},\ and\ \citenamefont {Kaminer}}]{pizzi2023light}%
  \BibitemOpen
  \bibfield  {author} {\bibinfo {author} {\bibfnamefont {A.}~\bibnamefont {Pizzi}}, \bibinfo {author} {\bibfnamefont {A.}~\bibnamefont {Gorlach}}, \bibinfo {author} {\bibfnamefont {N.}~\bibnamefont {Rivera}}, \bibinfo {author} {\bibfnamefont {A.}~\bibnamefont {Nunnenkamp}},\ and\ \bibinfo {author} {\bibfnamefont {I.}~\bibnamefont {Kaminer}},\ }\bibfield  {title} {\bibinfo {title} {Light emission from strongly driven many-body systems},\ }\href {https://www.nature.com/articles/s41567-022-01910-7} {\bibfield  {journal} {\bibinfo  {journal} {Nature Physics}\ }\textbf {\bibinfo {volume} {19}},\ \bibinfo {pages} {551} (\bibinfo {year} {2023})}\BibitemShut {NoStop}%
\bibitem [{\citenamefont {Lewenstein}\ and\ \citenamefont {Javanainen}(1987)}]{lewenstein1987cooperative}%
  \BibitemOpen
  \bibfield  {author} {\bibinfo {author} {\bibfnamefont {M.}~\bibnamefont {Lewenstein}}\ and\ \bibinfo {author} {\bibfnamefont {J.}~\bibnamefont {Javanainen}},\ }\bibfield  {title} {\bibinfo {title} {Cooperative quantum jumps with two atoms},\ }\href {https://journals.aps.org/prl/abstract/10.1103/PhysRevLett.59.1289} {\bibfield  {journal} {\bibinfo  {journal} {Physical review letters}\ }\textbf {\bibinfo {volume} {59}},\ \bibinfo {pages} {1289} (\bibinfo {year} {1987})}\BibitemShut {NoStop}%
\bibitem [{\citenamefont {Mandel}(1979)}]{mandel1979sub}%
  \BibitemOpen
  \bibfield  {author} {\bibinfo {author} {\bibfnamefont {L.}~\bibnamefont {Mandel}},\ }\bibfield  {title} {\bibinfo {title} {Sub-poissonian photon statistics in resonance fluorescence},\ }\href {https://opg.optica.org/ol/fulltext.cfm?uri=ol-4-7-205&id=6952} {\bibfield  {journal} {\bibinfo  {journal} {Optics letters}\ }\textbf {\bibinfo {volume} {4}},\ \bibinfo {pages} {205} (\bibinfo {year} {1979})}\BibitemShut {NoStop}%
\bibitem [{\citenamefont {Stammer}(2025{\natexlab{b}})}]{stammer2025high}%
  \BibitemOpen
  \bibfield  {author} {\bibinfo {author} {\bibfnamefont {P.}~\bibnamefont {Stammer}},\ }\bibfield  {title} {\bibinfo {title} {{High harmonic generation from a Bose-Einstein condensate}},\ }\href {https://arxiv.org/abs/2509.19022} {\bibfield  {journal} {\bibinfo  {journal} {arXiv:2509.19022}\ } (\bibinfo {year} {2025}{\natexlab{b}})}\BibitemShut {NoStop}%
\bibitem [{\citenamefont {Arg{\"u}ello-Luengo}\ \emph {et~al.}(2024)\citenamefont {Arg{\"u}ello-Luengo}, \citenamefont {Rivera-Dean}, \citenamefont {Stammer}, \citenamefont {Maxwell}, \citenamefont {Weld}, \citenamefont {Ciappina},\ and\ \citenamefont {Lewenstein}}]{arguello2024analog}%
  \BibitemOpen
  \bibfield  {author} {\bibinfo {author} {\bibfnamefont {J.}~\bibnamefont {Arg{\"u}ello-Luengo}}, \bibinfo {author} {\bibfnamefont {J.}~\bibnamefont {Rivera-Dean}}, \bibinfo {author} {\bibfnamefont {P.}~\bibnamefont {Stammer}}, \bibinfo {author} {\bibfnamefont {A.~S.}\ \bibnamefont {Maxwell}}, \bibinfo {author} {\bibfnamefont {D.~M.}\ \bibnamefont {Weld}}, \bibinfo {author} {\bibfnamefont {M.~F.}\ \bibnamefont {Ciappina}},\ and\ \bibinfo {author} {\bibfnamefont {M.}~\bibnamefont {Lewenstein}},\ }\bibfield  {title} {\bibinfo {title} {Analog simulation of high-harmonic generation in atoms},\ }\href {https://journals.aps.org/prxquantum/abstract/10.1103/PRXQuantum.5.010328} {\bibfield  {journal} {\bibinfo  {journal} {PRX Quantum}\ }\textbf {\bibinfo {volume} {5}},\ \bibinfo {pages} {010328} (\bibinfo {year} {2024})}\BibitemShut {NoStop}%
\bibitem [{\citenamefont {Arg{\"u}ello-Luengo}\ \emph {et~al.}(2025)\citenamefont {Arg{\"u}ello-Luengo}, \citenamefont {Rivera-Dean}, \citenamefont {Stammer}, \citenamefont {Ciappina},\ and\ \citenamefont {Lewenstein}}]{arguello2025quantum}%
  \BibitemOpen
  \bibfield  {author} {\bibinfo {author} {\bibfnamefont {J.}~\bibnamefont {Arg{\"u}ello-Luengo}}, \bibinfo {author} {\bibfnamefont {J.}~\bibnamefont {Rivera-Dean}}, \bibinfo {author} {\bibfnamefont {P.}~\bibnamefont {Stammer}}, \bibinfo {author} {\bibfnamefont {M.~F.}\ \bibnamefont {Ciappina}},\ and\ \bibinfo {author} {\bibfnamefont {M.}~\bibnamefont {Lewenstein}},\ }\bibfield  {title} {\bibinfo {title} {Quantum kramers-henneberger transformation},\ }\href {https://arxiv.org/abs/2507.13006} {\bibfield  {journal} {\bibinfo  {journal} {arXiv:2507.13006}\ } (\bibinfo {year} {2025})}\BibitemShut {NoStop}%
\bibitem [{\citenamefont {Bergquist}\ \emph {et~al.}(1986)\citenamefont {Bergquist}, \citenamefont {Hulet}, \citenamefont {Itano},\ and\ \citenamefont {Wineland}}]{bergquist1986observation}%
  \BibitemOpen
  \bibfield  {author} {\bibinfo {author} {\bibfnamefont {J.~C.}\ \bibnamefont {Bergquist}}, \bibinfo {author} {\bibfnamefont {R.~G.}\ \bibnamefont {Hulet}}, \bibinfo {author} {\bibfnamefont {W.~M.}\ \bibnamefont {Itano}},\ and\ \bibinfo {author} {\bibfnamefont {D.~J.}\ \bibnamefont {Wineland}},\ }\bibfield  {title} {\bibinfo {title} {Observation of quantum jumps in a single atom},\ }\href {https://journals.aps.org/prl/abstract/10.1103/PhysRevLett.57.1699} {\bibfield  {journal} {\bibinfo  {journal} {Physical Review Letters}\ }\textbf {\bibinfo {volume} {57}},\ \bibinfo {pages} {1699} (\bibinfo {year} {1986})}\BibitemShut {NoStop}%
\bibitem [{\citenamefont {Lewenstein}(2002)}]{lewenstein2002quantum}%
  \BibitemOpen
  \bibfield  {author} {\bibinfo {author} {\bibfnamefont {M.}~\bibnamefont {Lewenstein}},\ }\bibfield  {title} {\bibinfo {title} {Quantum jump statistics for two-atom systems},\ }\href {https://ieeexplore.ieee.org/document/980} {\bibfield  {journal} {\bibinfo  {journal} {IEEE journal of quantum electronics}\ }\textbf {\bibinfo {volume} {24}},\ \bibinfo {pages} {1403} (\bibinfo {year} {2002})}\BibitemShut {NoStop}%
\end{thebibliography}%

\appendix

\begin{center}
    \textbf{APPENDIX}
\end{center}

\section{\label{end:heisenberg}Details on the Heisenberg picture and first-order correlation function}

In this Appendix, we give a brief summary of the relevant results from the companion paper~\cite{stammer2025theory}, where we have provided the Heisenberg picture of quantum optical HHG, and introduced the quantum theory of optical coherence for the process. In particular, we discussed the first-order field correlation function
\begin{align}
\label{eq:first_order_correlation}
    G^{(1)}(t_1,t_2) = \expval*{E^{(-)}(t_1) E^{(+)}(t_2)},
\end{align}
where we considered the decomposition of the field operator $\vb{E}_Q (t) = \vb{E}^{(+)}(t) + \vb{E}^{(-)}(t) $, into a positive and negative field component 
\begin{align}
    \vb{E}^{(+)}(t) & =  i \hbar g \sum_{q} \sqrt{q} \pol_{q} a_{q}(t), \\
    \vb{E}^{(-)}(t) & = \left[\vb{E}^{(+)}(t) \right]^\dagger,
\end{align}
with the time-dependent Heisenberg operator $a_q(t)$. These are the solutions of the Heisenberg equations of motion 
\begin{align}
\label{eq:aq_heisenberg_eom}
    \dv{t} a_{q}(t) = - i \omega_{q} a_{q}(t) + g \sqrt{q} \sum_{i=1}^N \pol_{q} \cdot \vb{d}_i(t),
\end{align}
where $\vb{d}_i(t) = U^\dagger(t) \vb{d}_i(0) U(t) $ is the dipole moment operator in the respective Heisenberg picture.
Given the solution of Eq.~\eqref{eq:aq_heisenberg_eom}, as expressed in Eq.~\eqref{eq:aq_heisenberg_solution}, we can write the field operator as 
\begin{align}
    \vb{E}^{(+)}(t) = \vb{E}_f^{(+)}(t) + \vb{E}_s^{(+)}(t),
\end{align}
with 
\begin{align}
    \vb{E}_f^{(+)}(t) = i \hbar g \sum_{q} \sqrt{q} \pol_{q} a_{q} e^{- i \omega_{q} t},
\end{align}
and
\begin{equation}
\begin{aligned}
\label{eq:scattered_field}
    \vb{E}_s^{(+)}(t) &=  i \hbar g^2 \sum_{q} q \pol_{q} \int_{t_0}^t \dd t^\prime e^{- i \omega_{q} (t-t^\prime) } \\
    &\quad \times \sum_{i=1}^N \pol_{q} \cdot \vb{d}_i (t^\prime). 
\end{aligned}
\end{equation}

Here, $\vb{E}_f^{(+)}(t)$ describes the evolution of the free field in the absence of sources, and $\vb{E}_s^{(+)}(t)$ is the source field radiated by the atomic emitter.
Using the expression for the field operator in the definition of the first-order correlation function in Eq.~\eqref{eq:first_order_correlation}, we find that it can be decomposed into two terms
\begin{align}
    G^{(1)}(t,t+\tau) = G^{(1)}_{coh}(t,t+\tau) + G^{(1)}_{inc}(t,t+\tau).
\end{align}

This is obtained when decomposing the dipole moment correlations, appearing due to the expectation value over the atomic degrees of freedom of the source term in Eq.~\eqref{eq:scattered_field}, into a mean field contribution and its fluctuations
\begin{align}
\label{eq:dipole_single_atom}
    \expval{d(t_1) d(t_2)} = \expval{d(t_1)} \expval{d(t_2)} + \expval{\Delta d(t_1) \Delta d(t_2)},
\end{align}
where we have defined the fluctuations of the dipole around its mean $\Delta d(t) \equiv d(t) - \expval{d(t)}$. 
Hence, the coherent and incoherence contribution to the correlation function are respectively given by 
\begin{align}
    G^{(1)}_{coh}(t,t+\tau) & = \hbar^2 g^4 q^2 e^{- i \omega_q \tau} \int_{t_0}^t dt_1 \expval{d(t_1)} e^{-i \omega_q t_1} \nonumber \\
    & \quad \times \int_{t_0}^{t+\tau} dt_2  \expval{d(t_2)} e^{i \omega_q t_2}, \\
    G^{(1)}_{inc}(t,t+\tau) & = \hbar^2 g^4 q^2 e^{- i \omega_q \tau} \int_{t_0}^t dt_1 \, e^{-i \omega_q t_1} \\
    & \quad \times \int_{t_0}^{t+\tau} dt_2 \, e^{i \omega_q t_2} \expval{\Delta d(t_1) \Delta d(t_2)}. \nonumber
\end{align}

An evaluation of the first-order correlation functions can be found in the companion paper~\cite{stammer2025theory}, and in the Supplementary Material.

\section{\label{end:low_order}Second-order correlation function for low order harmonics and many emitters}

\begin{figure}
	\centering
	\includegraphics[width = 0.5\columnwidth]{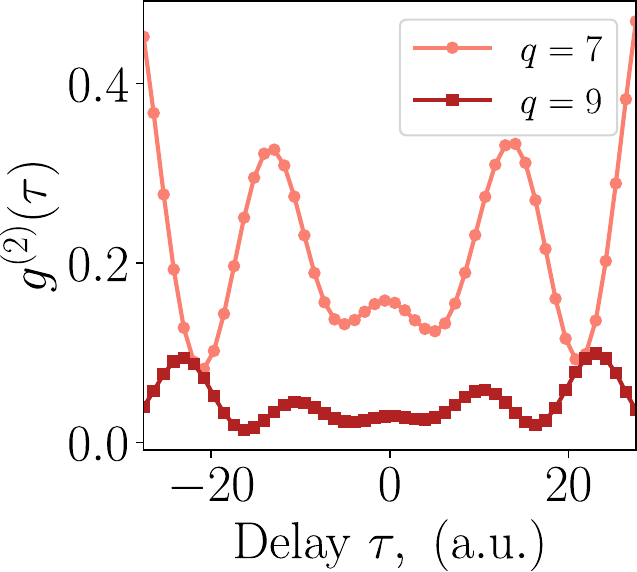}
	\caption{Normalized second-order correlation function $g^{(2)}(\tau)$ for low-order harmonics $q = \{7, \, 9 \}$. The time-delay $\tau$ is taken to be after one complete cycle of the driving laser field $T = 2 \pi / \omega$. The same field conditions as those in Fig.~\ref{fig:g2_function} are used. }
	\label{fig:low_order}
\end{figure}

\begin{figure}
    \centering
	\includegraphics[width = \columnwidth]{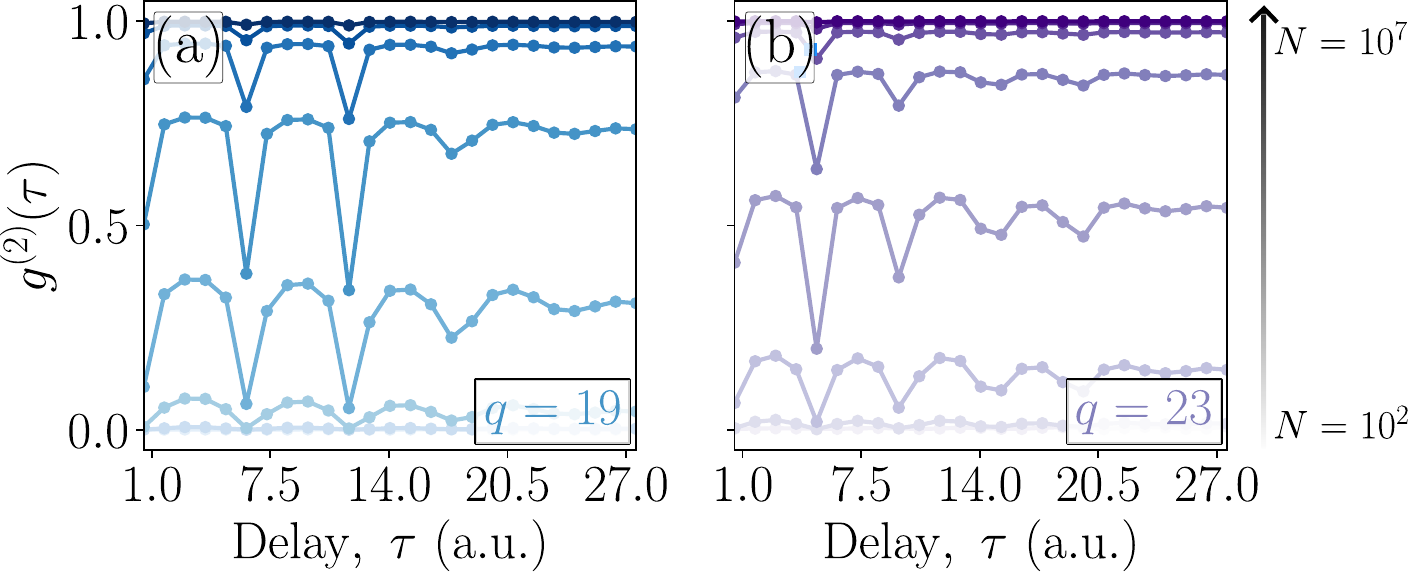}
	\caption{Normalized intensity correlation $g^{(2)}(\tau)$ for harmonic orders $q= 19$ in (a) and $q=23$ in (b) for varying number of emitters $N$, showing the same behavior as Fig~\ref{fig:dipole_higher_order}. The atom number increases for increasing color brightness including $\log_{10}N \in \{2.0, 2.7,3.4,4.1,4.8,5.5,6.7,7.0 \} $. The same field conditions as those in Fig.~\ref{fig:g2_function} are used.}
      \label{fig:g2_endmatter}
\end{figure}

In the main text, we have discussed the second-order correlation function in HHG, and in Fig.~\ref{fig:g2_function} we show the $g^{(2)}(\tau)$ function for the entire spectrum of non-perturbative harmonics starting from $q \ge 11$ until the harmonic cut-off at $q=23$. For completeness, we show the $g^{(2)}(\tau)$ also for the low-order harmonics in Fig.~\ref{fig:low_order}, indicating the same behavior and photon anti-bunching signature in HHG. 
Furthermore, in the main text we have shown the behavior of $g^{(2)}(\tau)$ for an increasing number of emitters for the $q= 13$ harmonic in Fig.~\ref{fig:dipole_higher_order}. For completeness, in Fig.~\ref{fig:g2_endmatter} we show the many emitter scenario for the harmonics $q=19$ and $q=23$.

\section{\label{end:many_atom}Many atom contribution}

In the following we show the derivation of the many-atom contributions to the higher-order dipole correlations in Eq.~\eqref{eq:dipole_high_order_expanded} of the main text.
We shall first look at the higher order dipole moment expectation value in Eq. \eqref{eq:dipole_correlation_2nd_order} for the $N$ atom case along the laser polarization direction
\begin{align}
\label{eq:2nd_order_dipole_correlation_2terms}
    \sum_{i,j,k,l=1}^N &  \bra{\vb{\bar g}} d_i(t_1) d_j(t_2) d_k(t_3) d_l(t_4) \ket{\vb{\bar g}} \nonumber \\
    & = \sum_{i=1}^N \expval{d_i(t_1) d_i(t_2) d_i(t_3) d_i(t_4)} \\
    & + \sum_{i \neq \{j,k,l\} } \sum_{j,k,l,=1}^N \expval{d_i(t_1) d_j(t_2) d_k(t_3) d_l(t_4)}, \nonumber
\end{align}
where $i \neq \{ j,k,l \}$ indicates that $i$ is not equal to $j,k,l$ at the same time.
The first term is the contribution from each of the $N$ emitters individually, and the second term takes into account the correlations between different emitters. We find that the first term scales as $\mathcal{O}(N)$, while the second term scales as $\mathcal{O}(N^4)$. 
Now, we consider the same assumptions of the main text which includes~\cite{stammer2025theory}: (i) an uncorrelated initial state of the emitters such that the dipole moment between different emitters factorize, and (ii) each emitter experiences the same driving field such that the subscript can be dropped due to the indistinguishability of the atoms. 
We thus have for the first term in Eq. \eqref{eq:2nd_order_dipole_correlation_2terms}
\begin{align}
    \sum_{i=1}^N \expval{d_i(t_1) d_i(t_2) d_i(t_3) d_i(t_4)} = N \expval{d(t_1) d(t_2) d(t_3) d(t_4)},
\end{align}
which is similar to the term evaluated in the single atom case with a linear scaling for the number of atoms $N$. 
For the second term, including the correlations between the emitters, we have 
\begin{align}
\label{eq:dipole_series_expansion}
    & \sum_{i \neq \{j,k,l\} }  \sum_{j,k,l,=1}^N \expval{d_i(t_1) d_j(t_2) d_k(t_3) d_l(t_4)} \\
    & = \frac{N!}{(N-4)!} \expval{d(t_1)} \expval{d(t_2)} \expval{d(t_3)} \expval{d(t_4)} + \mathcal{O}(N^3), \nonumber
\end{align}
where we have neglected terms of order $\mathcal{O}(N^3)$ since HHG experiments are typically performed in the $N\gg 1$ regime such that the major contribution comes from the coherent part. The coherent contribution, in which all dipole moment expectation values are evaluated independently, scales as $\mathcal{O}(N^4)$. In the Supplementary Material we show the next leading terms in the expansion.

\end{document}